\documentclass[11pt,a4paper]{article}

\usepackage{fullpage}
\usepackage{amsmath}
\usepackage{amssymb}
\usepackage{mathrsfs}
\usepackage{graphicx}
\usepackage{psfrag}
\usepackage[all,poly,curve,knot,cmtip]{xy}
\usepackage{citesort}

\newcommand {\be} {\begin{eqnarray*}}
\newcommand {\ee} {\end{eqnarray*}}
\newcommand {\bea} {\begin{eqnarray}}
\newcommand {\eea} {\end{eqnarray}}

\newcommand{\bm}[1]{\boldsymbol{#1}}

\newcommand{\ave}[1]{\langle {#1} \rangle}

\newcommand{\map}[3]{{#1}:{#2}\rightarrow{#3}}

\newcommand{\pdiff}[2]{\frac{\partial{#1}}{\partial{#2}}}

\newcommand{\tq}{\tilde{q}}
\newcommand{\tomega}{\tilde{\omega}}

\title{Extending the isolated horizon phase space to string-inspired gravity models}
\author{\textbf{Tom$\acute{\mbox{a}}\check{\mbox{s}}$ Liko}\footnote{A thesis presented
to Memorial University of Newfoundland in fulfillment of the thesis requirement for
the degree of Doctor of Philosophy in Theoretical Physics.  St. John's, Newfoundland,
Canada, 2008.  $\copyright$ Tom$\acute{\mbox{a}}\check{\mbox{s}}$ Liko 2008.}~\footnote{New
address: Institute for Gravitation and the Cosmos, Pennsylvania State University, University
Park, PA 16802, USA.}\\
\\{\small \it Department of Physics and Physical Oceanography}\\
{\small \it Memorial University of Newfoundland}\\
{\small \it St. John's, Newfoundland, Canada, A1B 3X7}}

\begin{document}

\maketitle

\begin{abstract}

An isolated horizon (IH) is a null hypersurface at which
the geometry is held fixed.  This generalizes the notion of
an event horizon so that the black hole is an object that
is in local equilibrium with its (possibly) dynamic
environment.  The first law of IH mechanics that arises
from the framework relates quantities that are all defined
\emph{at the horizon}.

IHs have been extensively studied in Einstein gravity
with various matter couplings and rotation, and in
asymptotically flat and asymptotically anti-de Sitter
(ADS) spacetimes in all dimensions $D\geq3$.  Motivated
by the nonuniqueness of black holes in higher dimensions
and by the black-hole/string correspondence principle,
we devote this thesis to the extension of the framework
to include IHs in string-inspired gravity models,
specifically to Einstein-Maxwell-Chern-Simons (EM-CS)
theory and to Einstein-Gauss-Bonnet (EGB) theory in
higher dimensions.  The focus is on determining the
generic features of black holes that are solutions to
the field equations of the theories under consideration.
To this end, we construct a covariant phase space for
both theories; this allows us to prove that the
corresponding weakly IHs (WIHs) satisfy the zeroth and
first laws of black-hole mechanics.

For EM-CS theory, we find that in the limit when the
surface gravity of the horizon goes to zero there is
a topological constraint.  Specifically, the integral
of the scalar curvature of the cross sections of the
horizon has to be positive when the dominant energy
condition is satisfied and the cosmological constant
$\Lambda$ is zero or positive.  There is no constraint
on the topology of the horizon cross sections when
$\Lambda<0$.  These results on topology of IHs are
independent of the material content of the
stress-energy tensor, and therefore the conclusions
for EM-CS theory carry over to theories with arbitrary
matter fields (minimally) coupled to Einstein gravity.

In addition, we consider rotating IHs in asymptotically
ADS and flat spacetimes, and find the restrictions that
are imposed on them if one assumes they are supersymmetric.
For the existence of a null Killing spinor in
four-dimensional $N=2$ gauged supergravity we show that
ADS supersymmetric isolated horizons (SIHs) are necessarily
extremal, that rotating SIHs must have non-trivial
electromagnetic fields, and that non-rotating SIHs
necessarily have constant curvature horizon cross sections
and a magnetic (though not electric) charge.  When the
cosmological constant is zero then the gravitational
angular momentum vanishes identically and the corresponding
SIHs are strictly non-rotating.  Likewise for the existence
of a null Killing spinor in five-dimensional $N=1$
supergravity, we show that SIHs (in asymptotically flat
spacetimes) are strictly non-rotating and extremal.

For EGB theory we restrict our study to non-rotating WIHs
and show explicitly that the expression for the entropy
appearing in the first law is in agreement with those
predicted by the Euclidean and Noether charge methods.
By carefully examining a concrete example of two
Schwarzschild black holes in a flat four-dimensional
spacetime that are merging, we find that the
area-increase law can be violated for certain values
of the GB parameter.  This provides a constraint on
the free parameter.

\end{abstract}

\section{Introduction}

\noindent{\emph{``What is mind?  No matter.  What is matter?  Never mind.''}}
$\sim$ H J Simpson

\subsection{Statement of the problem}

It has been appreciated for some time that a black hole behaves as a thermal object
and has a macroscopic entropy $\mathscr{S}_{\rm BH}$, the Bekenstein-Hawking entropy,
that is proportional to the surface area $\mathscr{A}$ of the event horizon
(Bekenstein 1973; Bekenstein 1974; Hawking 1975).  This fact is a very beautiful
example of the profound relationship between the classical and quantum aspects of
the gravitational field, and is one of the main reasons why the study of black holes
continues to be one of the most interesting areas of research in gravitational
theory.  It is also the main reason to believe that the gravitational field should
have a quantum description.  One of the goals of all the different approaches to
quantum gravity is to identify the microscopic degrees of freedom that account for
the entropy, and to obtain the area-entropy relation from first principles using
statistical mechanics.  If it turned out that gravity cannot be quantized, then
this fact would provide a very striking counterexample to our belief that thermal
properties \emph{of any object} are described quantum mechanically in terms of the
microstates of the corresponding system.

To get a general feeling for the problem, it is worth looking at the entropy with a
concrete example.  First, let us write down the expression with all physical constants.
For a black hole in \emph{Einstein gravity} this is
\bea
\mathscr{S}_{\rm BH} = \frac{\mathscr{A}k_{\rm B}c^{3}}{4\hbar G_{D}}
            = \frac{\mathscr{A}k_{\rm B}}{4l_{\rm P}^{2}} \, ,
\label{entropy1}
\eea
with $k_{\rm B}$ the Boltzmann constant and $l_{\rm P}^{2}$ the Planck ``area'' defined
by the speed of light $c$, $D$-dimensional gravitational constant $G_{D}$ and Planck
constant $\hbar$.  Let us further consider a Schwarzschild black hole of one solar mass
$M_{\odot}=1.989\times10^{30}$ $\mbox{kg}$ in four dimensions.  The spacetime for this
solution in spherical coordinates is the line element
\bea
dS^{2} = -\left(1-\frac{2M_{\odot}G_{4}}{rc^{2}}\right)c^{2}dt^{2}
         + \left(1-\frac{2M_{\odot}G_{4}}{rc^{2}}\right)^{-1}dr^{2}
         + r^{2}(d\theta^{2} + \sin^{2}\theta d\phi^{2}) \; .
\label{schwarzschild}
\eea
The event horizon has radius $r=2G_{4}M_{\odot}/c^{2}$ and the surface area is then
$\mathscr{A}=32\pi G_{4}^{2}M_{\odot}^{2}/c^{4}$.  This gives a numerical value of
\bea
\mathscr{S} = \frac{8\pi G_{4}k_{\rm B}M_{\odot}^{2}}{\hbar c} = 2.895 \times 10^{54}~
\mbox{J}\cdot\mbox{K}^{-1}
\label{entropy2}
\eea
for the entropy of the black hole.  The number of quantum states $\mathcal{N}$ that
this entropy corresponds to is therefore
\bea
\mathcal{N} = \mbox{exp}\left(\frac{\mathscr{S}}{k_{\rm B}}\right)
            = \mbox{exp}(2.098\times 10^{77}) \, ,
\eea
which is a huge number by any standards.  For comparison, we note that the number
$\mathscr{S}/k_{\rm B}$ is on the same order of magnitude as the estimated total
number of nucleons in the universe!

The problem is to answer the following question: \emph{What are the microscopic degrees
of freedom that account for the entropy of the black hole?}  The Schwarzschild solution
is static, which implies that the degrees of freedom cannot be gravitons.  They must be
described by nonperturbative configurations of the gravitational field.

The leading approaches to quantum gravity that have been most successfully applied
to the problem of black-hole microstates are loop quantum gravity (LQG) (Ashtekar
and Lewandowski 2004) and superstring theory (ST) (Aharoney \emph{et al} 2000).

\begin{itemize}

\item
\emph{Loop quantum gravity.}
Here one counts the states arising from punctures where spin networks traversally
intersect a surface that is specified in the quantized phase space with a set of
boundary conditions (Ashtekar \emph{et al} 1998; Ashtekar \emph{et al} 2000a).
This surface represents the black hole horizon and is intrinsically flat.
Curvature is induced at the punctures where the spin networks intersect the
surface and give it ``quanta of area''.  The LQG framework has been successful
in describing the statistical mechanics of all black holes (with simple topologies)
in four dimensions, but up to a single free parameter that enters into the classical
phase space as an ambiguity in the choice of real self-dual connection (Barbero 1996;
Immirzi 1997; Rovelli and Thiemann 1997).  In order for the framework to produce the
correct coefficient that matches the one for $\mathscr{S}$ in (\ref{entropy1}), this
parameter must be fixed to a specific value which depends on how the state-counting
is done.  See e.g. (Tamaki and Nomura 2005).  It was recently pointed out, however,
that if the Newton constant as well as the surface area of the black hole are
renormalized then the entropy may be the same for all values of the parameter
(Jacobson 2007).  Also, certain special properties of four-dimensional spacetimes
have to be exploited within the framework, which are crucial for the calculations
to work at all.  This makes it difficult to extend the framework to higher dimensions.

\item
\emph{Superstring theory.}
There are two (independent) approaches to the problem here.  The first is the D-brane
picture (Maldacena 1996; and references therein), whereby one counts the states of a
particular quantum field theory on a configuration of D-branes which forms a black
hole in the limit when the string coupling is increased.  The second is the anti-de
Sitter/conformal field theory (ADS/CFT) picture (Witten 1998a; 1998b), whereby a
black hole in a five-dimensional ADS spacetime is described by a conformally
invariant $SU(N)$ super Yang-Mills theory; here the states accounting for the
entropy are the quantum states of the CFT.  Both of these approaches have been
successful in describing the statistical mechanics, with the exact coefficient for
the area-entropy relation, but for a very limited class of black holes: extremal
and near-extremal in the D-brane picture while very small black holes (corresponding
to high temperature limit) in the ADS/CFT picture.  In particular, astrophysical
black holes such as those described by the solution (\ref{schwarzschild}) are not
among the class of black holes that are described in the ST approaches.

\end{itemize}

The LQG and ST approaches are very different, both philosophically and in the methods
that are used for quantization.  LQG on the one hand is a background independent
canonically quantized theory of pure gravity in four dimensions, while ST on the other
hand is a quantum field theory over a fixed nondynamical background in higher dimensions
that is supposed to describe all interactions as well as gravity.  It is unclear, and
surprising, that such different approaches all lead to the same answer.  This is an
instance of the ``problem of universality'' which has been advocated for some time now
by Carlip (2007).  Essentially, the entropy of a black hole may be fixed universally
by the diffeomorphism invariance of general relativity.

\subsection{Motivation for thesis research}

The fact that the ST approaches give the exact coefficient for the entropy of a
black hole is truly remarkable, despite that they do so for such a limited class
of solutions.  Nevertheless, the ST approaches are the most favored because they
explain the entropy \emph{dynamically}, and also from an aesthetically pleasing
point of view that ST is a unified theory of all interactions.  Despite all
successes though, a number of problems remain.  Among the more serious ones are
black hole nonuniqueness in higher dimensions and an inconsistency that has been
overlooked in the black-hole/string correspondence principle.

\begin{itemize}

\item
\emph{Black-hole nonuniqueness.}
In a four-dimensional asymptotically flat spacetime, a charged and rotating black
hole is uniquely described by its conserved charges; the only unique solution is
the Kerr-Newman metric (Robinson 1973).  This is a statement of the black-hole
uniqueness theorem, and is a striking property of the simplicity of black holes
in nature.  The advent of ST revolutionized our view of the universe, for example
with the requirement of extra spatial dimensions.  For a long time it was generally
assumed that the properties of four-dimensional black holes, particularly the
uniqueness theorem, simply carry over to higher dimensions as well.  The black ring
solution (Emparan and Reall 2002) that describes a rotating black hole with horizon
topology $S^{1}\times S^{2}$ in five dimensions, was the first counter-example to
the uniqueness of black holes in asymptotically flat spacetimes.  Specifically,
the uniqueness theorem fails (in five dimensions) because the conserved charges of
the ring can coincide with the conserved charges of a rotating black hole with
horizon topology $S^{3}$ (Myers and Perry 1986).  The natural question that should
be investigated is therefore the following: \emph{What properties of black holes
in four dimensions carry over to higher-dimensional spacetimes?}  More specifically,
we should ask the following question: \emph{What are the generic features of black
holes in higher-dimensional spacetimes in general and within the ST framework in
particular?}  An ideal method of investigating such questions is to employ a
covariant phase space of all solutions to the equations of motion for a given
action principle.

\item
\emph{Black-hole/string correspondence principle.}
The methods that are employed in ST lead to a first law of black-hole mechanics that
relates quantities at the event horizon and quantities defined at infinity.  This
``hybrid'' relation appears to be unphysical in ST from the point of view of the
black-hole/string correspondence principle (Susskind 1993), which states that there
is a smooth transition from a black hole to a string in the limit when the string
coupling is decreased.  For this correspondence principle to work, the entropies of
the black hole and string are required to be equal for a particular value of the string
coupling constant because the entropy of the black hole is proportional to the \emph{mass
squared} and the entropy of the string is proportional to the \emph{mass} (Horowitz and
Polchinski 1997).  However, while the mass of the black hole is measured at infinity,
the mass of the string is determined by the string coupling and tension which are
intrinsic quantities of the string state in the sense that no reference needs to be
made to infinity at all.  Therefore the conserved charges of the black hole state
should \emph{not} be defined at infinity.

\end{itemize}

The moral to be extracted from the above considerations is that a framework for black
holes in ST should be employed that is both quasilocal and general enough to allow for
a large class of solutions to be investigated.  Remarkably, such a framework does exist!
This is the isolated horizon (IH) framework (Ashtekar and Krishnan 2004).  The classical
theory of IHs was motivated by earlier considerations by Hayward (1994), but the framework
is considerably different as covariant phase space methods
(Witten 1986; Crnkovi$\acute{\mbox{c}}$ 1987; Crnkovi$\acute{\mbox{c}}$ and Witten 1987;
Crnkovi$\acute{\mbox{c}}$ 1988; Lee and Wald 1990; Ashtekar \emph{et al} 1991; Wald and
Zoupas 2000) are employed in the case of IHs.  All the quantities that appear in the first
law of IH mechanics are defined intrinsically at the horizon.  The concept of such a
surface generalizes the notion of a Killing horizon in stationary spacetimes to much more
general and therefore physical spacetimes that may include external radiation fields that
are dynamical.  Examples of such systems in general relativity are given by the so-called
Robertson-Trautman spacetimes (Ashtekar \emph{et al} 1999; Lewandowski 2000).

The IH framework may fit naturally into ST and the black-hole/string correspondence
principle.  The work presented here is a first step towards extending the IH phase space
beyond Einstein gravity so that a quasilocal description of black holes may be realized
within the context of ST.  In this thesis the framework is extended first to
Einstein-Maxwell-Chern-Simons (EM-CS) theory (Liko and Booth 2008; Booth and Liko 2008)
and then to Einstein-Gauss-Bonnet (EGB) theory (Liko and Booth 2007; Liko 2008) in higher
dimensions.  There are of course many more theories of gravity in higher dimensions.
Some of the modern approaches in five dimensions incorporating a large extra dimension
include braneworld cosmology (Maartens 2004) and induced-matter theory (Liko \emph{et al}
2004).

The motivation for extending the IH framework specifically to the two theories presented
in this thesis came from their relevance within the context of ST.  EM-CS theory is
important in ST because, in five dimensions, the corresponding action with negative
cosmological constant is the bosonic action of $N=1$ gauged supergravity (Cremmer 1980;
G$\ddot{\mbox{u}}$naydin \emph{et al} 1984; G$\ddot{\mbox{u}}$naydin \emph{et al} 1985);
black holes in particular are described by solutions to the bosonic equations of motion
with all fermionic fields and their variations vanishing in the vacuum (Gibbons \emph{et al}
1994; Gauntlett \emph{et al} 1999; Gutowski and Reall 2004).  In addition, the action
in four dimensions reduces to the Einstein-Maxwell (EM) action, which is the bosonic
action of $N=2$ gauged supergravity (Gibbons \emph{et al} 1994).  EGB theory is important
in ST because the corresponding action contains the only possible combination of
curvature-squared interactions for which the linearized equations of motion do not
contain any ghosts (Zwiebach 1985; Zumino 1986; Myers 1987).  This is particularly
important in ST because of the no-ghost theorem (Polchinski 1998), which states that
the BRST inner product is positive.

\subsection{Overview and main results}

In Section 2 we consider the phase space of solutions to the equations of motion
for the EM-CS action
\bea
S = \frac{1}{2\kappa_{D}}\int_{\mathcal{M}}d^{D-1}x\sqrt{-g}
      \Biggl\lbrace R
&-& 2\Lambda - \frac{1}{4}F^{2}\nonumber\\
&-& \frac{2\lambda}{3\sqrt{3}}
      \epsilon_{ab_{1}\cdots b_{D-1}}
      A^{a}F^{b_{1}b_{2}}\cdots F^{b_{D-2}b_{D-1}}\Biggr\rbrace \; .
\label{action1}
\eea
Here, $R$ is the scalar curvature, $g$ is the determinant of the spacetime metric
tensor $g_{ab}$ ($a,b,\ldots \in\{0,\ldots,D-1\}$), $A_{a}$ is the vector potential
and $F_{ab}=\partial_{a}A_{b}-\partial_{b}A_{a}$ (with $F^{2}=F_{ab}F^{ab}$) is the
field strength.  The constants appearing in the action are the gravitational coupling
constant $\kappa_{D}=8\pi G_{D}$ and the cosmological constant $\Lambda$.  The
cosmological constant is given by
\bea
\Lambda = \frac{\varepsilon}{2L^{2}}(D-1)(D-2) \, ,
\eea
where $\varepsilon\in\{-1,1\}$ and $L$ is the (anti-)de Sitter radius.  Also, we set
$c=\hbar=1$ from here on unless otherwise stated.  The last term is a Chern-Simons
(CS) term for the electromagnetic field; here $\lambda=0$ if $D$ is even and
$\lambda=1$ if $D$ is odd.  The field equations that are derived from the action
(\ref{action1}) when the metric is varied are the Einstein equations
\bea
G_{ab} = 2T_{ab} - \Lambda g_{ab}
\label{emcsfieldequations1}
\eea
with the Einstein tensor $G_{ab}$ and stress-energy tensor $T_{ab}$ given by
\bea
G_{ab} \equiv R_{ab} - \frac{1}{2}Rg_{ab}
\quad
\mbox{and}
\quad
T_{ab} = F_{ac}F_{b}^{\phantom{a}c} - \frac{1}{4}g_{ab}F^{2} \; .
\label{stressenergytensor}
\eea
The field equations that are derived from the action (\ref{action2}) when the vector
potential is varied are the Maxwell-Chern-Simons equations
\bea
\nabla_{\! a}F^{ab} = \frac{4(D+1)\lambda}{3\sqrt{3}\sqrt{-g}}\epsilon^{bc_{1}\cdots c_{D-1}}
                      F_{c_{1}c_{2}}\cdots F_{c_{D-2}c_{D-1}} \; .
\label{emcsfieldequations2}
\eea
There are several solutions to these equations that describe black holes.  In four dimensions
with $\lambda=0$ the equations are solved by a family of topological Kerr-Newman-ADS (KN-ADS)
spacetimes (Kosteleck$\acute{\mbox{y}}$ and Perry 1996; Caldarelli and Klemm 1999).  The
solutions that are supersymmetric describe: (a) rotating and extremal black holes with horizon
cross sections of spherical, cylindrical or toroidal topologies and having non-trivial
electromagnetic fields; and (b) non-rotating and extremal black holes with constant curvature
horizon cross sections of genus $g>1$ and with magnetic (but not electric) charge.  In five
dimensions, with $\Lambda=0$, the simplest solution is the five-dimensional
Reissner-Nordstr$\ddot{\mbox{o}}$m (RN) spacetime (Tangherlini 1963; Myers and Perry 1986).
The equations admit two asymptotically flat solutions that describe supersymmetric black holes.
These are the Breckenridge-Myers-Peet-Vafa (BMPV) black hole (Breckenridge \emph{et al} 1997),
and the Elvang-Emparan-Mateos-Reall (EEMR) black ring (Elvang \emph{et al} 2004).  The
Gutowski-Reall (GR) black hole is a generalization to ADS spacetime of the BMPV black hole
(Gutowski and Reall 2004).  The main purpose of the work in (Liko and Booth 2008; Booth and
Liko 2008) was to develop a quasilocal framework for these black holes.

First, we examine the boundary conditions and their consequences. To this end, we
consider the action (\ref{action1}) in the first-order connection formulation of
general relativity, after which we specify the boundary conditions that are imposed
onto the inner boundary of $\mathcal{M}$.  These boundary conditions capture the
notion of a weakly isolated horizon (WIH) that physically corresponds to an isolated
black hole in a surrounding spacetime with (possibly dynamical) fields and leads to
the zeroth law of black-hole mechanics.

Next we investigate the mechanics of the WIHs.  We show that the action principle with
boundaries is well defined by explicitly showing that the first variation of the surface
term vanishes on the horizon.  We then find an expression for the symplectic structure
by integrating over a spacelike $(D-1)$-surface the antisymmetrized second variation of
the surface term and adding to this the pullback of the resulting two-form to the WIH.
This allows us to find an expression for the local version of the (equilibrium) first
law of black-hole mechanics in dimensions $D\geq5$.

Summarizing thus far, we have the following:

\noindent{\bf Result 1.}
\emph{A charged and rotating WIH $\Delta\subset\mathcal{M}$ on the phase space
of solutions of EM-CS theory in $D$ dimensions satisfies the zeroth and first laws
of black-hole mechanics.}

After proving that the first law holds, we restrict our study to the stronger notion of
(fully) IHs.  These are WIHs for which the extrinsic as well as intrinsic geometries are
invariant under time translations.  For these horizons, the sign of the surface gravity
$\kappa_{(\ell)}$ is well defined.  The requirement that $\kappa_{(\ell)}\geq0$ therefore
allows us to define a parameter that provides a constraint on the topology of the IHs.
We find that the integral of the scalar curvature of the cross sections of the IH (in a
spacetime with nonnegative cosmological constant) have to be strictly positive if the
dominant energy condition is satified.  Furthermore, this integral will be zero if the
horizon is extremal and non-rotating, and the stress-energy tensor $T_{ab}$ is of the
form such that $T_{ab}\ell^{a}n^{b}=0$ for any two null vectors $\ell$ and $n$ with
normalization $\ell_{a}n^{a}=-1$ at the horizon.  For negative cosmological constant
there is no restriction on the scalar curvature of the cross sections of the IH.

Summarizing now, we have the following:

\noindent{\bf Result 2.}
\emph{The IH cross sections in a higher-dimensional spacetime with nonnegative cosmological
constant are of positive Yamabe type; if $\Lambda<0$ then there is no restriction on the
sign of the scalar curvature.}

This result is in agreement with recent work on the topological constraints of
higher-dimensional black holes in globally stationary spacetimes (Helfgott \emph{et al}
2006; Galloway 2006; Galloway and Schoen 2006).  We note that the physical content of the
stress-energy tensor at this point is completely arbitrary.  Therefore Result 2 implies
that the topology considerations are valid for any matter (nonminimally) coupled to
Einstein gravity.  In the case of electromagnetic fields with or without the CS term,
the scalar $T_{ab}\ell^{a}n^{b}$ is the square of the electric flux crossing the horizon.

In Section 3 we examine the restrictions that are imposed on IHs if one assumes that they are
supersymmetric.  To do this we specialize to IHs in four dimensions with negative cosmological
constant and in five dimensions with vanishing cosmological constant.  The former theory is
the bosonic part of four-dimensional $N=2$ gauged supergravity, and the latter theory is the
bosonic part of five-dimensional $N=1$ supergravity.  We show that the existence of a Killing
spinor in four dimensions requires that the induced (normal) connection $\omega$ on the horizon
has to be non-zero unless the electric charge (but not magnetic charge) vanishes, and that the
surface gravity $\kappa$ has to be zero.  The former condition means that the gravitational
component of the horizon angular momentum is non-zero provided that $\omega$ is not a closed
one-form.  The latter condition means that the IH is extremal.  Likewise, we show that the
existence of a Killing spinor in five dimensions requires that $\omega$ vanishes and this
immediately also gives $\kappa=0$.

Summarizing now, we have the following:

\noindent{\bf Result 3.}
\emph{A SIH of four-dimensional $N=2$ gauged supergravity is extremal, and is
either: (a) rotating with non-trivial electromagnetic field; or (b) non-rotating with
constant curvature horizon cross sections and magnetic (but not electric) charge.  A SIH
of four-dimensional $N=2$ supergravity and of five-dimensional $N=1$ supergravity with
zero cosmological constant is non-rotating and extremal.}

The topological KN-ADS family of solutions (Caldarelli and Klemm 1999) describe rotating and
extremal supersymmetric black holes in four-dimensional ADS spacetime.  Among the special cases
is a solution describing a non-rotating and extremal black hole with constant curvature
horizon cross sections and magnetic charge.  The BMPV solution describes an extremal black
hole with nonvanishing angular momentum and non-rotating Killing horizon; this black hole
solution is an example of a distorted IH with arbitrary rotations in the bulk fields
(Ashtekar \emph{et al} 2004).  When the angular momentum vanishes this solution reduces
to the extremal RN solution in isotropic coordinates.  The conclusions drawn from our Result
2 together with Result 3 are that the only possible horizon topologies for SIHs are $S^{2}$ in
four dimensions (when $\Lambda=0$) or $S^{3}$ and $S^{1}\times S^{2}$ in five dimensions.
Both these topologies have been realized and the corresponding solutions, for example the BMPV
black hole (Breckenridge \emph{et al} 1997) and the EEMR black ring (Elvang \emph{et al} 2004),
are well known.  The torus topology is a special case that can occur only if the stress-energy
tensor is of the form such that $T_{ab}\ell^{a}n^{b}=0$ for any two null vectors $\ell$ and $n$
with normalization $\ell_{a}n^{a}=-1$.  A solution describing such a black hole has yet to be
discovered.

In Section 4 we consider the phase space of solutions to the equations of motion for the
EGB action
\bea
S = \frac{1}{2\kappa_{D}}\int_{\mathcal{M}}d^{D}x\sqrt{-g}
    \left[R - 2\Lambda + \alpha\left(R^{2}-4R_{ab}R^{ab}+R_{abcd}R^{abcd}\right)\right] \; .
\label{action2}
\eea
In addition to the quantities that also appear in the EM-CS action (\ref{action1}),
the action (\ref{action2}) also contains explicit dependence on the Riemann tensor
$R_{abcd}$ and Ricci tensor $R_{ab}=R_{\phantom{a}acb}^{c}$.  The constant $\alpha$
is the GB parameter.  The field equations that are derived from the action (\ref{action2})
when the metric is varied are the EGB equations
\bea
G_{ab} = -\Lambda g_{ab} &+& \alpha\Biggl\lbrace\frac{1}{2}\left(R^{2} - 4R_{cd}R^{cd}
                          + R_{cdef}R^{cdef}\right)g_{ab}\nonumber\\
                         &-& 2RR_{ab} + 4R_{ac}R_{b}^{\phantom{a}c} + 4R_{acbd}R^{cd}
                          - 2R_{acde}R_{b}^{\phantom{a}cde}\Biggr\rbrace \; .
\label{egbfieldequations}
\eea
When $\alpha=0$ these equations reduce to the vacuum Einstein equations.  There are
several solutions to these equations that describe black holes.  The simplest were
found independently by Boulware and Deser (1985) and by Wheeler (1986a; 1986b), and
describe a static spherically symmetric black hole.  The causal structure and
thermodynamics of this solution were later studied by Myers and Simon (1988).  The
solution was subsequently extended to ADS spacetime by Cai (2002) and by Cho and
Neupane (2002).  The purpose of the work in (Liko and Booth 2007; Liko 2008) was to
develop a quasilocal framework for these black holes.

Just as for IHs in EM-CS theory, we begin by examining the boundary conditions and
their consequences.  It turns out that for the zeroth law to be satisfied, the
boundary conditions need to be slightly modified.  Specifically, an analogue of
the dominant energy condition has to be imposed onto the Ricci tensor instead of the
matter stress-energy tensor.  The zeroth law then follows naturally from the modified
boundary conditions.

Next we investigate the mechanics of the WIHs.  In particular, we show that the action
principle for WIHs in EGB theory is well defined by explicitly showing that the first
variation of the surface term vanishes on the horizon.  This turns out to be quite
complicated due to the presence of the GB term.  Nevertheless, we verify the
differentiability of the action for EGB theory by brute force at the expense of
restricting the phase space to non-rotating WIHs.  We then find an expression for the
symplectic structure by integrating over a spacelike $(D-1)$-surface the antisymmetrized
second variation of the surface term and adding to this the pullback of the resulting
two-form to the WIH.  This allows us to find an expression for the local version of the
(equilibrium) first law of black-hole mechanics in dimensions $D\geq5$, with an entropy
expression that contains a correction term that is proportional to the surface integral
of the scalar curvature of the cross sections of the horizon.  We demonstrate the validity
of our expression for the quasilocal entropy of WIHs by directly comparing it to those
expressions that are obtained by the Euclidean (Cai 2002; Cho and Neupane 2002) and
Noether charge (Clunan \emph{et al} 2004) methods.

Summarizing thus far, we have the following:

\noindent{\bf Result 4.}
\emph{A non-rotating WIH $\Delta\subset\mathcal{M}$ on the phase space of solutions
of EGB theory in $D$ dimensions satisfies the zeroth and first laws of black-hole
mechanics.}

We conclude our investigation of IHs in EGB theory by looking at physical consequences
of the correction term in the entropy can have on the area-increase law.  In order to
make the analysis concrete, the calculation is done for black holes in four dimensions,
specifically for the merging of two Schwarzschild black holes in flat spacetime.  It
turns out that for this very special case the second law of black-hole mechanics will
be violated if $\alpha$ is greater than the product of the masses of the black holes
before merging minus a small correction due to radiation that may be lost by
gravitational waves during the merging process.

Summarizing now, we have the following:

\noindent{\bf Result 5.}
\emph{There is a lower bound on $\alpha$ for which the area-increase law will be violated
when two black holes merge.}

The calculation of the bound on $\alpha$ is done in four dimensions.  However, a similar
bound may presumably be derived for specific solutions in higher dimensions as well
[although in this case the topologies are not as severely restricted as they are in four
dimensions, even for Einstein gravity with $\Lambda=0$ (Helfgott \emph{et al} 2006;
Galloway 2006; Galloway and Schoen 2006)].  Result 2 also corrects a long-held
misconception about the GB term, namely that its presence in four dimensions does not
lead to any physical effects because the term is a topological invariant and does not
show up in the equations of motion.

In Section 5 we conclude the thesis with a brief summary of the work that has been done
here, and discuss some classical applications of IHs in EM-CS theory and EGB theory.

\section{Isolated Horizons in EM-CS Theory}

\noindent{\emph{``The beginner ... should not be discouraged if ... he finds
that he does not have the prerequisite for reading the prerequisites.''}}
$\sim$ P Halmos

\subsection{First-order action for EM-CS theory}

For application to IHs, we work with the ``connection-dynamics'' formulation of general
relativity.  For details we refer the reader to the review (Ashtekar and Lewandowski
2004) and references therein.  In this formulation, the configuration space consists of
the triple $(e^{I},A_{\phantom{a}J}^{I},\bm{A})$; the coframe
$e^{I}=e_{a}^{\phantom{a}I}dx^{a}$ $(I,J,\ldots\in\{0,\ldots,D-1\})$ determines the
spacetime metric
\bea
g_{ab} = \eta_{IJ}e_{a}^{\phantom{a}I} \otimes e_{b}^{\phantom{a}J} \, ,
\eea
the gravitational ($SO(D-1,1)$) connection
$A_{\phantom{a}J}^{I}=A_{a\phantom{a}J}^{\phantom{a}I}dx^{a}$ determines
the curvature two-form
\bea
\Omega_{\phantom{a}J}^{I}
= dA_{\phantom{a}J}^{I} + A_{\phantom{a}K}^{I} \wedge A_{\phantom{a}J}^{K} \, ,
\eea
and the electromagnetic ($U(1)$) connection $\bm{A}$ determines the curvature
\bea
\bm{F} = d\bm{A} \; .
\eea
In this thesis, spacetime indices $a,b,\ldots$ are raised and lowered using the metric
$g_{ab}$, while internal Lorentz indices $I,J,\ldots$ are raised and lowered using the
Minkowski metric $\eta_{IJ}=\mbox{diag}(-1,1,\ldots,1)$.  The curvature $\Omega$ defines
the Riemann tensor $R_{\phantom{a}JKL}^{I}$ [with the convention of Wald (1984)] via
\bea
\Omega_{\phantom{a}J}^{I}
= \frac{1}{2}R_{\phantom{a}JKL}^{I}e^{K} \wedge e^{L} \; .
\eea
The Ricci tensor is then $R_{IJ}=R_{\phantom{a}IKJ}^{K}$, and the
Ricci scalar is $R=\eta^{IJ}R_{IJ}$.  The gauge covariant derivative
$\mathscr{D}$ acts on generic fields $\Psi_{IJ}$ such that
\bea
\mathscr{D}\Psi_{\phantom{a}J}^{I}
= d\Psi_{\phantom{a}J}^{I}
  + A_{\phantom{a}K}^{I} \wedge \Psi_{\phantom{a}J}^{K}
  - A_{\phantom{a}J}^{K} \wedge \Psi_{\phantom{a}K}^{I} \; .
\eea
The coframe defines the $(D-m)$-form
\bea
\Sigma_{I_{1}\ldots I_{m}}
= \frac{1}{(D-m)!}\epsilon_{I_{1} \ldots I_{m}I_{m+1} \ldots I_{D}}
e^{I_{m+1}} \wedge \cdots \wedge e^{I_{D}} \, ,
\eea
where the totally antisymmetric Levi-Civita tensor $\epsilon_{I_{1} \ldots I_{D}}$
is related to the spacetime volume element by
\bea
\epsilon_{a_{1} \ldots a_{D}}=\epsilon_{I_{1} \ldots I_{D}}
e_{a_{1}}^{\phantom{a}I_{1}} \cdots e_{a_{D}}^{\phantom{a}I_{D}} \; .
\eea

In this configuration space, the action (\ref{action1}) for EM-CS theory
on the manifold $(\mathcal{M},g_{ab})$ (assumed for the moment to have
no boundaries) is given by
\bea
S = \frac{1}{2\kappa_{D}}\int_{\mathcal{M}}\Sigma_{IJ} \wedge \Omega^{IJ}
    - 2\Lambda\bm{\epsilon}
    - \frac{1}{4}\bm{F} \wedge \star \bm{F} - \frac{2\lambda}{3\sqrt{3}}\bm{A}
      \wedge \bm{F}^{(D-1)/2} \; .
\label{action5}
\eea
Here $\bm{\epsilon}=e^{0} \wedge \cdots \wedge e^{D-1}$ is the spacetime
volume element and ``$\star$'' denotes the Hodge dual.

The equations of motion are given by $\delta S=0$, where $\delta$ is
the first variation; i.e. the stationary points of the action.  For this
configuration space the equations of motion are derived from independently
varying the action with respect to the fields $(e,A,\bm{A})$.  To get the
equation of motion for the coframe we note the identity
\bea
\delta\Sigma_{I_{1} \ldots I_{m}}
= \delta e^{M} \wedge \Sigma_{I_{1} \ldots I_{m}M} \; .
\eea
This leads to
\bea
\Sigma_{IJK} \wedge \Omega^{JK} - 2\Lambda\Sigma_{I}
= \mathscr{T}_{I} \, ,
\label{eom1}
\eea
where $\mathscr{T}_{I}$ denotes the electromagnetic stress-energy ($D-1$)-form.
The equation of motion for the connection $A$ is
\bea
\mathscr{D}\Sigma_{IJ} = 0 \, ;
\label{eom2}
\eea
this equation says that the torsion $T^{I}=\mathscr{D}e^{I}$ is zero.  The
equation of motion for the connection $\bm{A}$ is
\bea
d \star \bm{F} - \frac{4(D+1)\lambda}{3\sqrt{3}}\bm{F}^{(D-1)/2} = 0 \; .
\label{eom3}
\eea
The second term in this equation is the contribution due to the CS term in the action.
In even dimensions the equation reduces to the standard Maxwell equation $d \star \bm{F}=0$.
The equations (\ref{eom1}) and (\ref{eom2}) are equivalent to the field equations
(\ref{emcsfieldequations1}) and (\ref{emcsfieldequations2}) in the metric formulation,
with the components of $\mathscr{T}_{I}$ identified with the electromagnetic stress-energy
tensor.

\subsection{Boundary conditions}

Let us from here on consider the manifold $(\mathcal{M},g_{ab})$ to contain
boundaries; the conditions that we will impose on the inner boundary will capture
the notion of an isolated black hole that is in local equilibrium with its (possibly)
dynamic surroundings.  We follow the general recipe that was developed in (Ashtekar
\emph{et al} 2000c).

First we give some general comments about the structure of the manifold.  Specifically,
$\mathcal{M}$ is a $D$-dimensional Lorentzian manifold with topology $R\times M$,
contains a $(D-1)$-dimensional null surface $\Delta$ as inner boundary (representing
the horizon), and is bounded by $(D-1)$-dimensional spacelike manifolds $M^{\pm}$ that
extend from $\Delta$ to infinity.  The topology of $\Delta$ is
$R\times\mathbb{S}^{D-2}$, with $\mathbb{S}^{D-2}$ a compact ($D-2$)-space.  $M$ is a
partial Cauchy surface such that $\mathbb{S}^{D-2}\cong\Delta\cap M$.  See Figure 1.
\begin{figure}[t]
\begin{center}
\psfrag{D}{$\Delta$}
\psfrag{Mp}{$M^{+}$}
\psfrag{Mm}{$M^{-}$}
\psfrag{Mi}{$M$}
%\psfrag{Sp}{$\mathbb{S}^{+}$}
%\psfrag{Sm}{$\mathbb{S}^{-}$}
\psfrag{Sp}{$\phantom{Sp}$}
\psfrag{Sm}{$\phantom{Sm}$}
\psfrag{S}{$\mathbb{S}$}
\psfrag{B}{$\mathscr{B}$}
\psfrag{M}{$\mathcal{M}$}
\includegraphics[width=4.5in]{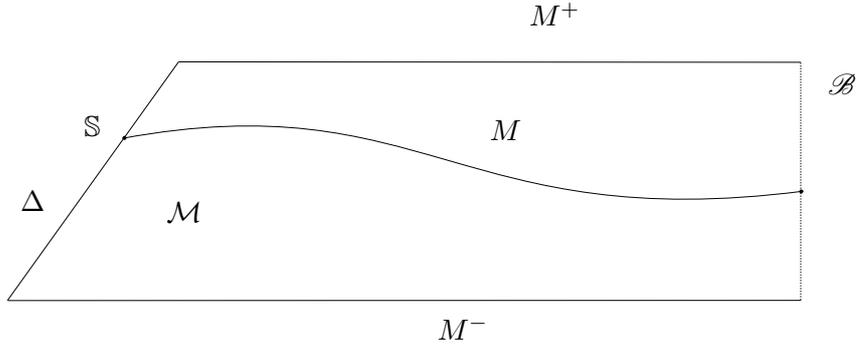}
\caption[The spacetime manifold $\mathcal{M}$ and its boundaries.]{The
spacetime manifold $\mathcal{M}$ and its boundaries.  The region of the
$D$-dimensional spacetime $\mathcal{M}$ being considered has an internal
boundary $\Delta$ representing the event horizon, and is bounded by two
$(D-1)$-dimensional spacelike hypersurfaces $M^{\pm}$ which extend from
the inner boundary $\Delta$ to the boundary at infinity $\mathscr{B}$.
$M$ is a partial Cauchy surface that intersects $\Delta$ in a compact
($D-2$)-space $\mathbb{S}$.}
\end{center}
\end{figure}

The outer boundary $\mathscr{B}$ is some arbitrary $(D-1)$-dimensional surface.  With the
exception of $\S 2.6$, we consider the purely quasilocal case in this chapter and neglect
any subleties that are associated with the outer boundary.  Including this contribution in
the phase space amounts to imposing fall-off conditions on the fields for fixed $\Lambda$
[e.g. asymptotically flat (Ashtekar \emph{et al} 2000c) or asymptotically ADS (Ashtekar
\emph{et al} 2007)] as they approach $\mathscr{B}$.  In $\S 2.6$ we briefly discuss
rotation in asymptotically ADS spacetimes.

$\Delta$ is a WIH, which is defined in the following way:

\noindent{\bf Definition I.}
\emph{A WIH $\Delta$ is a null surface
and has a degenerate metric $q_{ab}$ with signature $0+\ldots+$ (with $D-2$
non-degenerate spatial directions) along with an equivalence class of null normals $[\ell]$
(defined by $\ell \sim \ell^\prime \Leftrightarrow \ell' = z \ell$ for some constant $z$) such that the
following conditions hold: (a) the expansion $\theta_{(\ell)}$ of $\ell_{a}$
vanishes on $\Delta$; (b) the field equations hold on $\Delta$; (c) the
stress-energy tensor is such that the vector $-T_{\phantom{a}b}^{a}\ell^{b}$ is a
future-directed and causal vector; (d) $\pounds_{\ell}\omega_{a}=0$ and
$\pounds_{\ell}\underleftarrow{\bm{A}}=0$ for all $\ell\in[\ell]$ (see below).}

The first three conditions determine the intrinsic geometry of $\Delta$. Since $\ell$ is normal to 
$\Delta$ the associated null congruence is necessarily twist-free and geodesic. By condition
(a) that congruence is non-expanding. Then the Raychaudhuri equation implies that
$T_{ab}\ell^{a}\ell^{b}=-\sigma_{ab}\sigma^{ab}$, with $\sigma_{ab}$ the shear tensor, and 
applying the energy condition (c) we find that $\sigma_{ab} = 0$. Thus, together these conditions
tell us that the intrinsic geometry of $\Delta$ is 
``time-independent'' in the sense that all of its (two-dimensional) cross sections have identical intrinsic geometries. 

Next,  the vanishing of the expansion, twist and shear imply that (Ashtekar \emph{et al} 2000c)
\bea
\nabla_{\!\underleftarrow{a}}\ell_{b}\approx\omega_{a}\ell_{b} \, ,
\label{connectionondelta}
\eea
with ``$\approx$'' denoting equality restricted to $\Delta$ and the underarrow
indicating pull-back to $\Delta$.  Thus the one-form $\omega$ is the natural connection
(in the normal bundle) induced on the horizon.  These conditions also imply that (Ashtekar \emph{et al} 2000c)
\bea
\underleftarrow{\ell \lrcorner \bm{F}} = 0 \; .
\label{pullback1}
\eea
With the field equations (\ref{eom3}) and the Bianchi identity $d\bm{F}=0$, it then
follows that
\bea
\pounds_{\ell}\underleftarrow{\bm{F}}
\approx \ell \lrcorner \underleftarrow{d\bm{F}}
          + d(\underleftarrow{\ell \lrcorner \bm{F}}) = 0 \; .
\label{liepullback}
\eea
This implies that the electric charge is independent of the choice of cross
sections $\mathbb{S}^{D-2}$ (Ashtekar \emph{et al} 2000b). Similarly (in four dimensions) the magnetic
charge is also a constant.

From (\ref{connectionondelta}) we find that 
\bea
\ell^{a}\nabla_{\! a}\ell^{b} = (\ell \lrcorner \omega) \ell^{b} \; ,
\eea
and define the surface gravity $\kappa_{(\ell)}=\ell \lrcorner \omega$ as the inaffinity of this geodesic
congruence. Note that it is certainly dependent on specific element of $[ \ell ]$ as under the transformation
$\ell \rightarrow z \ell$:
 \bea
%\omega^{\prime}_{a} = \omega_{a}
%\quad
%\mbox{and}
%\quad
\kappa_{(\ell)} \rightarrow  z \kappa_{(\ell)} \; .
\eea
In addition to the surface gravity, we also define the electromagnetic scalar
potential $\Phi_{(\ell)}=-\ell \lrcorner \bm{A}$ for each $\ell\in[\ell]$ and this has a similar dependence. 

Now, it turns out that if the first three conditions hold, then one can always find an 
equivalence class $[\ell]$ such that (d) also holds. Hence this last condition does not further
restrict the geometries under discussion, but only the scalings of the null normal. However, 
making such a choice ensures that (Ashtekar \emph{et al} 2000c):
\bea
d\kappa_{(\ell)} = d(\ell \lrcorner \omega) = 0
\quad
\mbox{and}
\quad
d\Phi_{(\ell)} = d(\ell \lrcorner \bm{A}) = 0 \; .
\eea
These conditions follow from the Cartan identity, (\ref{pullback1}) and the property that $d\omega$
is proportional to $\bm{\tilde{\epsilon}}$ [defined below in (\ref{areaelement})] (Ashtekar \emph{et
al} 2000c).  This establishes the zeroth law of WIH mechanics: the surface gravity and scalar
potential are constant on $\Delta$.

\subsection{Variation of the boundary term}

Let us now look at the variation of the action (\ref{action2}).  Denoting the
triple $(e,A,\bm{A})$ collectively as a generic field variable $\Psi$, the
first variation gives
\bea
\delta S = \frac{1}{2\kappa_{D}}\int_{\mathcal{M}}E[\Psi]\delta\Psi
           - \frac{1}{2\kappa_{D}}\int_{\partial\mathcal{M}}J[\Psi,\delta\Psi] \; .
\label{first}
\eea
Here $E[\Psi]=0$ symbolically denotes the equations of motion and
\bea
J[\Psi,\delta\Psi] = \Sigma_{IJ} \wedge \delta A^{IJ} - \bm{\Phi} \wedge \delta\bm{A}
\label{surface}
\eea
is the surface term with $(D-2)$-form
\bea
\bm{\Phi} = \star\bm{F} - \frac{4(D+1)\lambda}{3\sqrt{3}}\bm{A} \wedge \bm{F}^{(D-3)/2} \; .
\label{boldphi}
\eea
If the integral of $J$ on the boundary $\partial\mathcal{M}$ vanishes then the action
principle is said to be differentiable.  We must show that this is the case.  Because
the fields are held fixed at $M^{\pm}$ and at $\mathscr{B}$, $J$ vanishes there.
Therefore it suffices to show that $J$ vanishes at the inner boundary $\Delta$.  To
show that this is true we need to find an expression for $J$ in terms of $\Sigma$, $A$
and $\bm{A}$ pulled back to $\Delta$.  As for the gravitational variables, this is
accomplished by fixing an internal basis consisting of the (null) pair $(\ell,n)$ and
$D-2$ spacelike vectors $\vartheta_{(i)}$ ($i\in\{2,\ldots,D-1\}$) such that
\bea
e_{0} = \ell \, ,
\quad
e_{1} = n \, ,
\quad
e_{i} = \vartheta_{(i)} \, ,
\label{npbasis}
\eea
together with the conditions
\bea
\ell \cdot n=-1 \, ,
\quad
\ell \cdot \ell = n \cdot n = \ell \cdot \vartheta_{(i)} = n \cdot \vartheta_{(i)} = 0 \, ,
\quad
\vartheta_{(i)} \cdot \vartheta_{(j)} = \delta_{ij} \; .
\label{npconditions}
\eea
This basis represents a higher-dimensional analogue of the Newman-Penrose (NP) formalism
(Pravda \emph{et al} 2004).  The coframe $e_{a}^{\phantom{a}I}$ can be decomposed in terms
of the vectors in the basis (\ref{npbasis}) such that
\bea
e_{a}^{\phantom{a}I} = -\ell^{I}n_{a} - \ell_{a}n^{I}
                                        + \vartheta_{(i)}^{\phantom{a}I}\vartheta^{(i)}_{a} \, ;
\label{coframedecomp}
\eea
summation is understood over repeated spacelike indices ($i,j,k$ etc).  The pullback of the
coframe to $\Delta$ is therefore
\bea
e_{\underleftarrow{a}}^{\phantom{a}I}
\approx -\ell^{I}n_{a} + \vartheta_{(i)}^{\phantom{a}I}\vartheta^{(i)}_{a} \, ,
\label{pullbackofcoframe}
\eea
whence the $(D-2)$-form
\bea
\underleftarrow{\Sigma}_{IJ} &\approx&
-\frac{1}{(D-3)!}
\epsilon_{IJA_1 \dots A_{D-2}} \ell^{A_1} 
\vartheta^{\phantom{a}A_2}_{(i_1)} \dots \vartheta^{\phantom{a}A_{D-2}}_{(i_{D-3})}
\left(
n \wedge \vartheta^{(i_1)} \wedge \dots \wedge \vartheta^{(i_{D-3})} \right) 
\nonumber \\
& &
+ \frac{1}{(D-2)!} \epsilon_{IJA_1 \dots A_{D-2}}  
\vartheta^{\phantom{a}A_1}_{(i_1)} \dots \vartheta^{\phantom{a}A_{D-2}}_{(i_{D-2})}
\left( \vartheta^{(i_1)} \wedge \dots \wedge \vartheta^{(i_{D-2})} \right) \; .\nonumber\\
\label{pullbackofsigmaij}
\eea
To find the pull-back of $A$ we first note that
\bea
\nabla_{\!\underleftarrow{a}}\ell_{I}
&\approx& \nabla_{\!\underleftarrow{a}}\left(e_{\phantom{a}I}^{b}\ell_{b}\right)\nonumber\\
&\approx& (\nabla_{\!\underleftarrow{a}}e_{\phantom{a}I}^{b})\ell_{b}
              + e_{\phantom{a}I}^{b}\nabla_{\!\underleftarrow{a}}\ell_{b}\nonumber\\
&\approx& e_{\phantom{a}I}^{b}\omega_{a}\ell_{b}\nonumber\\
&\approx& \omega_{a}\ell_{I} \, ,
\label{pullback2}
\eea
where we used $\nabla_{\! a}e_{\phantom{a}I}^{b}=0$ in going from the
second to the third line (a consequence of the metric compatibility of
the connection). Then, taking the covariant derivative of $\ell$ acting on internal indices gives
\bea
\nabla_{\! a}\ell_{I} = \partial_{a}\ell_{I} + A_{aIJ}\ell^{J} \, ,
\eea
with $\partial$ representing a flat derivative operator that is compatible with
the internal coframe on $\Delta$.  Thus $\partial_{a}\ell_{I}\approx0$ and
\bea
\nabla_{\!\underleftarrow{a}}\ell_{I}\approx A_{\underleftarrow{a}IJ}\ell^{J} \; .
\eea
Putting this together with (\ref{pullback2}) we have that
\bea
A_{\underleftarrow{a}IJ}\ell^{J}\approx\omega_{a}\ell_{I} \, ,
\eea
and this implies that the pull-back of $A$ to the horizon is of the form
\bea
A_{\underleftarrow{a}}^{\phantom{a}IJ}
\approx -2\ell^{[I}n^{J]}\omega_{a}
+ a_{a}^{(i)}\ell^{[I}\vartheta_{(i)}^{\phantom{a}J]}
+ b_a^{(ij)}\vartheta_{(i)}^{\phantom{a}[I}\vartheta_{(j)}^{\phantom{a}J]} \, ,
\label{pullbackofa}
\eea
where the $a_{a}^{(i)}$ and $b_{a}^{(ij)}$ are one-forms in the cotangent
space $T^{*}(\Delta)$.  It follows that the variation of (\ref{pullbackofa})
is
\bea
\delta A_{\underleftarrow{a}}^{\phantom{a}IJ}
\approx -2\ell^{[I}n^{J]}\delta \omega_{a}
+ \delta a_{a}^{(i)}\ell^{[I}\vartheta_{(i)}^{\phantom{a}J]}
+ \delta b_a^{(ij)} \vartheta_{(i)}^{\phantom{a}[I}\vartheta_{(j)}^{J]} \; .
\label{variationofpullbackofa}
\eea
Finally, by direct calculation, it can be shown that the gravitational part
$J_{\rm Grav}$ of the surface term (\ref{surface}) reduces to
\bea
J_{\rm Grav}[\Psi,\delta\Psi]
\approx \bm{\tilde{\epsilon}} \wedge \delta \omega \; .
\label{simplifiedpullbackofcurrent}
\eea
Here,
\bea
\bm{\tilde{\epsilon}}=\vartheta^{(1)} \wedge \dots \wedge \vartheta^{(D-2)}
\label{areaelement}
\eea
is the area element of the cross sections $\mathbb{S}^{D-2}$ of the horizon.

Now we make use of the fact that, because $\ell$ is normal to the surface, its variation
will also be normal to the surface.  That is, $\delta\ell\propto\ell$ for some $\ell$
fixed in $[\ell]$.  This together with $\pounds_{\ell}\omega=0$ then implies that
$\pounds_{\ell}\delta\omega=0$.  However, $\omega$ is held fixed on $M^{\pm}$ which
means that $\delta\omega=0$ on the initial and final cross-sections of $\Delta$
(i.e. on $M^{-}\cap\Delta$ and on $M^{+}\cap\Delta$), and because $\delta\omega$ is
Lie dragged on $\Delta$ it follows that $J_{\rm Grav}\approx0$.  The same argument
also holds for the electromagnetic part $J_{\rm EM}$ of the surface term (\ref{surface}).
In particular, because the electromagnetic field is in a gauge adapted to the horizon,
$\pounds_{\ell}\underleftarrow{\bm{A}}=0$, and with $\delta\ell\propto\ell$ we also have
that $\pounds_{\ell}\delta\underleftarrow{\bm{A}}=0$.  This is sufficient to show that
$J_{\rm EM}\approx0$ as well.  Therefore the surface term $J|_{\partial\mathcal{M}}=0$
for the Einstein-Maxwell theory with electromagnetic CS term, and we conclude that the
equations of motion $E[\Psi]=0$ follow from the action principle $\delta S=0$.

\subsection{Covariant phase space}

The derivation of the first law involves two steps.  First we need to find the symplectic
structure on the covariant phase space $\bm{\Gamma}$ consisting of solutions $(e,A,\bm{A})$
to the field equations (\ref{eom1}), (\ref{eom2}) and (\ref{eom3}) on $\mathcal{M}$.  Once
we have a suitable (closed and conserved) symplectic two-form, we then need to specify an
evolution vector field $\xi^{a}$.  In this section we derive the symplectic two-form.  In
the next section we will specify the evolution vector field which will also serve to
introduce an appropriate notion of horizon angular momentum.

The antisymmetrized second variation of the surface term gives the symplectic current,
and integrating over a spacelike hypersurface $M$ gives the symplectic structure
$\bm{\Omega}\equiv\bm{\Omega}(\delta_{1},\delta_{2})$ (with the choice of $M$ being
arbitrary).  Following (Ashtekar \emph{et al} 2000c), we find that the second variation
of the surface term (\ref{surface}) gives
\bea
J[\Psi,\delta_{1}\Psi,\delta_{2}\Psi]
= \delta_{1}\Sigma_{IJ} \wedge \delta_{2}A^{IJ}
  - \delta_{2}\Sigma_{IJ} \wedge \delta_{1}A^{IJ}
  - \delta_{1}\bm{\Phi} \wedge \delta_{2}\bm{A}
  - \delta_{2}\bm{\Phi} \wedge \delta_{1}\bm{A} \; .\nonumber\\
\eea
Whence integrating over $M$ defines the \emph{bulk} symplectic structure
\bea
\bm{\Omega}_{\rm B}
= \frac{1}{2\kappa_{D}}\int_{M}
  \left[\delta_{1}\Sigma_{IJ} \wedge \delta_{2}A^{IJ}
  - \delta_{2}\Sigma_{IJ} \wedge \delta_{1}A^{IJ}
  - \delta_{1}\bm{\Phi} \wedge \delta_{2}\bm{A}
  + \delta_{2}\bm{\Phi} \wedge \delta_{1}\bm{A}\right] \; .\nonumber\\
\label{bulksymplectic}
\eea
We also need to find the pull-back of $J$ to $\Delta$ and add the integral of this
term to $\bm{\Omega}_{\rm B}$ so that the resulting symplectic structure on
$\bm{\Gamma}$ is conserved.  If we define potentials $\psi$ and $\chi$ for the
surface gravity $\kappa_{(\ell)}$ and electric potential $\Phi_{(\ell)}$ such that
\bea
\pounds_{\ell}\psi \approx \ell \lrcorner \omega = \kappa_{(\ell)}
\quad
\mbox{and}
\quad
\pounds_{\ell}\chi \approx \ell \lrcorner \bm{A} = -\Phi_{(\ell)} \, ,
\label{potentials}
\eea
then the pullback to $\Delta$ of the symplectic structure will be a total
derivative; using the Stokes theorem this term becomes an integral over the cross
sections $\mathbb{S}^{D-2}$ of $\Delta$.  Hence the full symplectic structure is given by
\bea
\bm{\Omega}
&=& \frac{1}{2\kappa_{D}}\int_{M}
    \left[\delta_{1}\Sigma_{IJ} \wedge \delta_{2}A^{IJ}
    - \delta_{2}\Sigma_{IJ} \wedge \delta_{1}A^{IJ}
    - \delta_{1}\bm{\Phi} \wedge \delta_{2}\bm{A}
    + \delta_{2}\bm{\Phi} \wedge \delta_{1}\bm{A}\right]\nonumber\\
& &
    + \frac{1}{\kappa_{D}}\oint_{\mathbb{S}^{D-2}}
    \left[\delta_{1}\bm{\tilde{\epsilon}} \wedge \delta_{2}\psi
    - \delta_{2}\bm{\tilde{\epsilon}} \wedge \delta_{1}\psi
    + \delta_{1}\bm{\Phi} \wedge \delta_{2}\chi
    - \delta_{2}\bm{\Phi} \wedge \delta_{1}\chi\right] \; .\nonumber\\
\label{fullsymplectic}
\eea

\subsection{Angular momentum and the first law}

In $D$ dimensions, there are $\lfloor (D-1)/2 \rfloor$ rotation parameters given by the
Casimir invariants of the rotation group $SO(D-1)$.  Here, ``$\lfloor \cdot \rfloor$''
denotes the ``integer value of''.  For a multidimensional WIH rotating with
angular velocities $\Omega_{\iota}$ ($\iota=1,\ldots,\lfloor (D-1)/2 \rfloor$), a suitable
evolution vector field on the covariant phase space is given by (Ashtekar \emph{et al}
2001; Ashtekar \emph{et al} 2007)
\bea
\xi^{a} = z\ell^{a} + \sum_{\iota=1}^{\lfloor (D-1)/2 \rfloor}\Omega_{\iota}\phi_{\iota}^{a} \; .
\eea
Here, $\phi_{\iota}^{a}$ are spacelike rotational vector fields that satisfy
\bea
\pounds_{\phi}q_{ab}=0 \, ,
\quad
\pounds_{\phi}\ell_{a}=0 \, ,
\quad
\pounds_{\phi}\omega_{a}=0 \, ,
\quad
\pounds_{\phi}\underleftarrow{\bm{A}}=0 \, ,
\quad
\pounds_{\phi}\underleftarrow{\bm{F}}=0 \; .
\eea
The vector field $\xi$ is similar to the linear combination
$\zeta=t+\sum_{\iota}\Omega_{\iota}m_{\iota}$ (with $t$ a timelike Killing vector and
$m_{\iota}$ spacelike Killing vectors) for the KN solution.  By contrast, we note that
$\xi$ is spacelike in general and becomes null when all angular momenta are zero, while
$\zeta$ is null in general and becomes timelike when all angular momenta are zero.

Moving on, the first law now follows directly from evaluating the symplectic
structure at $(\delta,\delta_{\xi})$ (Ashtekar \emph{et al} 2007).  This gives two
surface terms: one at infinity (which is identified with the ADM energy), and one at
the horizon.  We find that the surface term at the horizon is given by
\bea
\bm{\Omega}|_{\Delta}
= \frac{1}{\kappa_{D}}\oint_{\mathbb{S}^{D-2}}\kappa_{(z\ell)}\delta\bm{\tilde{\epsilon}}
&+& \frac{1}{\kappa_{D}}\oint_{\mathbb{S}^{D-2}}\Phi_{(z\ell)}\delta\bm{\Phi}\nonumber\\
&+& \sum_{\iota=1}^{\lfloor (D-1)/2 \rfloor}\frac{\Omega_{\iota}}{\kappa_{D}}
  \delta\oint_{\mathbb{S}^{D-2}}\left[(\phi_{\iota} \lrcorner \omega)\bm{\tilde{\epsilon}}
  + (\phi_{\iota} \lrcorner \bm{A})\bm{\Phi}\right] \, ,
\label{evaluation}
\eea
where we used $\kappa_{(z\ell)}=\pounds_{z\ell}\psi=z\ell\lrcorner\omega$ and
$\Phi_{(z\ell)}=\pounds_{z\ell}\chi=z\ell\lrcorner\bm{A}$.  These potentials are constant
for any given horizon, but in general vary across the phase space from one point to another.
This implies that (\ref{evaluation}) is \emph{not} in general a total variation.  However,
if $\kappa_{(z\ell)}$, $\Phi_{(z\ell)}$ and $\Omega_{\iota}$ can be expressed as functions
of the entropy $\mathcal{S}$, charge $\mathcal{Q}$ and angular momenta $\mathcal{J}_{\iota}$
defined by
\bea
\mathcal{S}     &=& \frac{1}{4G_{D}}\oint_{\mathbb{S}^{D-2}}\bm{\tilde{\epsilon}}
\label{entropy}\\
\mathcal{Q}     &=& \frac{1}{8\pi G_{D}}\oint_{\mathbb{S}^{D-2}}\bm{\Phi}
\label{charge}\\
\mathcal{J}_{\iota} &=& \frac{1}{8\pi G_{D}}\oint_{\mathbb{S}^{D-2}}
                    \left[(\phi_{\iota} \lrcorner \omega)\bm{\tilde{\epsilon}}
                    + (\phi_{\iota} \lrcorner \bm{A})\bm{\Phi}\right] \,
\label{angularmomentum}
\eea
and satisfy the integrability conditions
\bea
\pdiff{\kappa}{\mathcal{J}} = \pdiff{\Omega}{\mathcal{S}} \, ,
\quad
\pdiff{\kappa}{\mathcal{Q}} = \pdiff{\Phi}{\mathcal{S}} \, ,
\quad
\pdiff{\Omega}{\mathcal{Q}} = \pdiff{\Phi}{\mathcal{J}} \, ,
\eea
then there exists a function $\mathcal{E}$ such that (Ashtekar \emph{et al} 2001;
Ashtekar \emph{et al} 2007)
\bea
\bm{\Omega}|_{\Delta}(\delta,\delta_{\xi})=\delta \mathcal{E} \; .
\eea
In this case (\ref{evaluation}) becomes
\bea
\delta \mathcal{E} = \frac{\kappa_{(z\ell)}}{2\pi}\delta \mathcal{S} + \Phi_{(z\ell)}\delta \mathcal{Q}
                     + \sum_{\iota=1}^{\lfloor (D-1)/2 \rfloor}
                       \Omega_{\iota}\delta \mathcal{J}_{\iota} \, ,
\label{firstlaw}
\eea
which is the first law (for a quasi-static process).  Therefore WIHs in $D$-dimensional EM-CS
theory satisfy the first law (and the zeroth law) of black-hole mechanics.  This is in agreement
with (Gauntlett \emph{et al} 1999), but with a very important difference.  Here, all the quantities
appearing in the first law are defined at the horizon; no reference was made to the boundary at infinity.

\noindent{\bf Remarks.} Several remarks are in order here.
\begin{enumerate}
\item
The expression (\ref{angularmomentum}) implies that the horizon angular momentum contains
contributions from both gravitational \emph{and} electromagnetic fields, here referred to as
$\mathcal{J}_{\rm Grav}$ and $\mathcal{J}_{\rm EM}$.  This is in contrast to the standard
angular momentum expressions at infinity, such as the Komar expression.  One can show
(Ashtekar \emph{et al} 2001; Ashtekar \emph{et al} 2007) that $\mathcal{J}_{\rm Grav}$ is
equivalent to the (quasilocal) Komar integral
\bea
\mathcal{J}_{\rm K} = -\frac{1}{8\pi G_{D}}\oint_{\mathbb{S}^{D-2}}\star d\phi \, ,
\eea
and this matches the expression for Killing horizons at infinity.
\item
It would appear that if $\mathcal{J}_{\rm Grav}=0$ then there is still a non-zero contribution
to (\ref{angularmomentum}) from $\mathcal{J}_{\rm EM}$.  However, it can be shown (Ashtekar
\emph{et al} 2001) that if $\phi$ is the restriction to $\Delta$ of a \emph{global} rotational
Killing field $\varphi$ contained in $\mathcal{M}$, then $\mathcal{J}_{\rm EM}$ is actually
the angular momentum of the electromagnetic radiation in the bulk.  What happens is that the
bulk integral $\smallint_{M}T_{ab}\varphi^{a}dS^{b}$ can be written as the sum of a surface
term at $\Delta$ and a surface term at $\mathscr{B}$.  Therefore we say that a non-rotating
WIH is one for which $\mathcal{J}_{\rm Grav}=0$.
\item
The charges at $\mathscr{B}$ are the charges of the spacetime and are independent of the
charges at $\Delta$.
\end{enumerate}

\subsection{Rotation in ADS spacetime}

Currently there is a lot of interest in the ADS/CFT correspondence (Maldacena 1998; Witten
1998a; Witten 1998b; Aharony \emph{et al} 2000).  A significant amount of effort on the
gravity side has been focused on finding charged and rotating black hole solutions in
five-dimensional ADS spacetime, both non-extremal in general (Hawking \emph{et al} 1999;
Chamblin \emph{et al} 1999; Hawking and Reall 1999; Klemm and Sabra 2001;
Cveti$\check{\mbox{c}}$ \emph{et al} 2004; Chong \emph{et al} 2005) and supersymmetric in
particular (Gutowski \emph{et al} 2004; Kunduri \emph{et al} 2006; Kunduri \emph{et al}
2007a; Kunduri and Lucietti 2007).

For these black holes, however, there is an ambiguity in how the conserved charges
are defined.  This was first pointed out by Caldarelli \emph{et al} (2000).  The
ambiguity arises because for rotating black holes in ADS spacetime there are two
distinct natural choices for the timelike Killing field.  To see this, let's
compare the KN solution (with $\Lambda=0$) and the KN-ADS solution.  The KN solution
contains the vector $K=\partial/\partial t$ with which one can define the charges.
When $\Lambda<0$, however, there is another timelike Killing vector in addition to $K$
that appears and is given by
$K^{\prime}=\partial/\partial t+(a/L^{2})\partial/\partial\phi$.  For the KN-ADS
solution, $K$ remains timelike \emph{everywhere} outside the event horizon which implies
that if $K$ is chosen as the generator of time translations then there is no ergoregion
present.  By contrast, $K^{\prime}$ becomes spacelike near the event horizon which
implies that if $K^{\prime}$ is chosen as the generator of time translations then there
is an ergoregion in the neighbourhood of the event horizon.  Physically, this means that
defining the conserved charges with respect to $K$ corresponds to a frame at infinity
that is co-rotating, whereas defining the conserved charges with respect to $K^{\prime}$
corresponds to a frame at infinity that is non-rotating.

The original motivation for defining the conserved charges with respect to $K$ was that
the corresponding boundary CFT conserved charges satisfy the first law of thermodynamics
(Hawking \emph{et al} 1999); but this comes at the cost that the bulk conserved charges
do not (Caldarelli \emph{et al} 2000; Gibbons \emph{et al} 2005).  This claim has by now
been corrected.  As was shown in (Gibbons \emph{et al} 2006), one can always pass from
the bulk conserved charges to the boundary conserved charges in such a way that both
sets separately satisfy the first law.  The key to this resolution is that the conserved
charges of a rotating black hole in ADS spacetime have to be measured with respect to the
timelike vector which corresponds to a frame that is non-rotating at infinity.

From the above considerations, it is clear that rotation in ADS spacetime should
be independent of the coordinates that are used.  This is especially crucial when
considering supersymmetric black holes in ADS spacetime (the extremal limit of a
non-rotating ADS black hole results in a naked singularity).  In this section we
will briefly discuss how the IH framework provides a resolution to the above pathology.

To begin, we define an asymptotically ADS spacetime.  Following (Ashtekar and Das 2000;
Ashtekar \emph{et al} 2007), we have the following:

\noindent{\bf Definition II.}
\emph{A spacetime $(\mathcal{M},g_{ab})$ is said to be asymptotically ADS if there exists
a spacetime $(\widehat{\mathcal{M}},\tilde{g}_{ab})$ with outer boundary $\mathscr{I}$ such
that $\widehat{\mathcal{M}}-\mathscr{I}$ is diffeomorphic to $\mathcal{M}$ and the following
conditions hold: (a) there exists a function $\Omega$ on $\widehat{\mathcal{M}}$ for which
$\tilde{g}_{ab}=\Omega^{2}g_{ab}$ on $\mathcal{M}$; (b) $\Omega$ vanishes on $\mathscr{I}$
but the gradient $\nabla_{a}\Omega$ is nowhere vanishing on $\mathscr{I}$; (c) the stress-energy
tensor $T_{ab}$ on $\mathcal{M}$ is such that $\Omega^{-(D-2)}T_{ab}$ has a smooth limit to
$\mathscr{I}$; and (d) the Weyl tensor $\tilde{C}_{abcd}$ of $\tilde{g}_{ab}$ is such that
$\Omega^{-(D-4)}\tilde{C}_{abcd}$ is smooth on $\mathcal{M}$ and vanishes on $\mathscr{I}$.}

These are the standard boundary conditions which have been tailored to ensure that a spacetime
will be asymptotically ADS.  Their meaning is discussed in detail in (Ashtekar and Das 2000).

In the presence of a negative cosmological constant and with no matter fields, the covariant
phase space of WIHs is modified to include a set of conserved charges at $\mathscr{I}$
(Ashtekar \emph{et al} 2007).  These are the Ashtekar-Magnon-Das (AMD) charges (Ashtekar and
Magnon 1984; Ashtekar and Das 2000)
\bea
\mathscr{Q}_{\xi}^{(\mathscr{I})} = \frac{L}{8\pi G_{D}}\oint_{\mathbb{C}^{D-2}}
                                    \widetilde{E}_{ab}k^{a}\tilde{u}^{b}\bm{\tilde{\varepsilon}} \, ,
\label{amdcharges}
\eea
with $k^{a}$ a Killing vector field that generates a symmetry (i.e. time translation etc),
$\tilde{u}^{a}$ the unit timelike normal to $\mathbb{C}^{D-2}$, $\bm{\tilde{\varepsilon}}$
the area form on $\mathbb{C}^{D-2}$ and $\widetilde{E}_{ab}$ the leading-order electric part
of the Weyl tensor.  Explicitly we have that
\bea
\widetilde{E}_{ab} = \frac{1}{D-3}\Omega^{3-D}\widetilde{C}_{abcd}\tilde{n}^{c}\tilde{n}^{d} \, ,
\eea
where $\tilde{n}^{a}=\tilde{\nabla}^{a}\Omega$.  As was shown in Appendix B of (Hollands
\emph{et al} 2005), inclusion of antisymmetric tensor fields in the action does not contribute
anything to the charges at $\mathscr{I}$ because the fields fall off too quickly.  Therefore
the charges at infinity for EM-CS theory are precisely the AMD charges (\ref{amdcharges}).

Gibbons \emph{et al} (2005) showed that the asymptotic time translation Killing field for
an exact solution has to be chosen in such a way that the frame at infinity is non-rotating.
If this is done then the AMD charge evaluated for the solution will result in an expression
for mass that satisfies the first law.  Moreover, Gibbons \emph{et al} (2006) showed that
using this definition for the asymptotic time translation has to be used for a consistent
transition to the conserved charges of the boundary CFT.

Let us summarize.  The IH framework provides a coherent physical picture whereby two sets
of conserved charges arise in ADS spacetime: the charges measured at infinity and the local
charges measured at the horizon.  The local conserved charges at the horizon then satisfy
the first law.  When evaluated on exact solutions to the field equations, the charges at
infinity correspond to asymptotic symmetries that are measured with respect to a
\emph{non-rotating} frame at infinity.

The description of ADS black holes presented here is somewhat different from the description
of black holes in globally stationary spacetimes where an ambiguity appears that manifests
itself as a choice of whether the conserved charges are measured with respect to a frame at
infinity that is rotating or non-rotating.  This ambiguity does not appear in the IH framework
essentially because the conserved charges of the black hole are measured \emph{at the horizon},
and the corresponding first law is intrinsic to the horizon with no mixture of quantities there
and at infinity!

\subsection{A topological constraint from extremality}

One of the properties of an extremal black hole is that its surface gravity is zero.
Another property is that its horizons are degenerate: the inner and outer horizons coincide.
As a result, an extremal black hole is one for which there are no trapped surfaces ``just
inside'' the horizon.  This property was recently used (Booth and Fairhurst 2008) to define
an extremality condition for quasilocal horizons.  We note here the evolution equation for
the expansion of the null normal $n^{a}$ (Booth and Fairhurst 2008):
\bea
\pounds_{\ell}\theta_{(n)} + \kappa \theta_{(n)} + \frac{1}{2}\mathcal{R}
= d_{a}\tilde{\omega}^{a} + \|\tilde{\omega}\|^{2} + (T_{ab}
  - \Lambda g_{ab})\ell^{a}n^{b} \; .
\label{nevolution}
\eea
Here, $\mathcal{R}$ is the scalar curvature of $\mathbb{S}^{D-2}$, $d_{a}$ is the
covariant derivative operator that is compatible with the metric
\bea
\tilde{q}_{ab} = g_{ab} + \ell_{a}n_{b} + \ell_{b}n_{a}
\label{dminustwometric}
\eea
on $\mathbb{S}^{D-2}$, and $\|\tilde{\omega}\|^{2}=\tilde{\omega}_{a}\tilde{\omega}^{a}$
where
%$\tilde{\omega}$
%is the projection of $\omega$ onto $\mathbb{S}^{D-2}$.  The relation between $\omega$
%and $\tilde{\omega}$ can be obtained by noting that from (\ref{connectionondelta})
%follows
%\bea
%\omega_{a}=-n^{b}\nabla_{a}\ell_{b} \, ,
%\eea
%and projecting this onto $\mathbb{S}^{D-2}$ using $\tilde{q}_{ab}$ gives:
\bea
\tilde{\omega}_{a} = \tilde{q}_{a}^{\phantom{a}b}\omega_{b}
%                  &=& -(g_{a}^{\phantom{a}b} + \ell_{a}n^{b} + \ell^{b}n_{a})
%                      n^{c}\nabla_{b}\ell_{c}\nonumber\\
                   = \omega_{a} + \kappa_{(\ell)}n_{a}
\label{amoneform}
\eea
is the projection of $\omega$ onto $\mathbb{S}^{D-2}$.  $\tilde{\omega}$ is referred
to as the rotation one-form.

Our desire is to apply the expression (\ref{nevolution}) to \emph{black holes}, and in order to
do this we need to impose some restrictions on the WIHs.  In order to proceed we now restrict our
attention to fully isolated horizons (IHs).  These are WIHs for which there is a scaling of the
null normals for which the commutator $[\pounds_{\ell},\mathcal{D}]=0$, where $\mathcal{D}$ is the
intrinsic covariant derivative on the horizon.  This means that not only is condition (d) of
Definition I satisfied, but also it implies that $[\pounds_{\ell},\mathcal{D}]n^{a}=0$ (Ashtekar
\emph{et al} 2002).  In contrast to the condition (d) for WIHs, then, this stronger condition
cannot always be met and geometrically such horizons not only have time-invariant intrinsic
geometry, they also have time-invariant extrinsic geometry.  That said it is clear that this
condition similarly fixes $\ell$ only up to a constant scaling. As such it does not uniquely
determine the value of the surface gravity $\kappa_{(\ell)}$ but does fix its sign. In
particular this allows us to invariantly say whether or not $\kappa_{(\ell)}$ vanishes. This
then gives rise to an invariant characterization of extremality that is intrinsic to the
horizon: a horizon is sub-extremal if $\kappa>0$ ($\theta_{(n)}<0$) and extremal (with
degenerate horizons) if $\kappa=0$.  Further, $\pounds_{\ell}\theta_{(n)}=0$ and combining
this with the fact that the inward expansion $\theta_{(n)}$ should always be less than zero,
an integration of (\ref{nevolution}) gives
\bea
\eta \equiv \oint_{\mathbb{S}^{D-2}}\bm{\tilde{\epsilon}}(T_{ab}\ell^{a}n^{b}
+ \Lambda + \|\tilde{\omega}\|^{2})
- \frac{1}{2}\oint_{\mathbb{S}^{D-2}}\bm{\tilde{\epsilon}}\mathcal{R} \leq 0 \; .
\label{characterization}
\eea
(Here we used $-\Lambda g_{ab}\ell^{a}n^{b}=\Lambda$ and the fact that
$\oint_{\mathbb{S}^{D-2}}\bm{\tilde{\epsilon}}d_{a}\tilde{\omega}^{a}=0$.)  This
inequality provides an alternative characterization of extremal IHs: if $\eta<0$
($\kappa>0$ and $\theta_{(n)}<0$) then $\Delta$ is nonextremal, and if $\eta=0$
($\kappa=0$) then $\Delta$ is extremal.  However, this inequality also provides a
topological constraint on the cross sections of $\Delta$.  To see this, rewrite
(\ref{characterization}) so that
\bea
\oint_{\mathbb{S}^{D-2}}\bm{\tilde{\epsilon}}(\mathcal{R} - 2\Lambda)
\geq 2\oint_{\mathbb{S}^{D-2}}\bm{\tilde{\epsilon}}(T_{ab}\ell^{a}n^{b}
+ \|\tilde{\omega}\|^{2}) \; .
\label{curvaturecondition}
\eea
Now, observe that the dominant energy condition requires that
$T_{ab}\ell^{a}n^{b}\geq0$.  In addition, $\|\tilde{\omega}\|^{2}$ is manifestly
non-negative.  The inequality (\ref{curvaturecondition}) therefore restricts the
topology of the cross sections of the horizon.  The condition (\ref{curvaturecondition})
is the same as the one that was found in four dimensions for marginally trapped surfaces
(Hayward 1994), nonexpanding horizons (Pawlowski \emph{et al} 2004) and dynamical horizons
(Ashtekar and Krishnan 2003; Booth and Fairhurst 2007).

For nonextremal horizons, $\eta<0$, and the constraint (\ref{curvaturecondition})
splits into two possibilities, depending on the nature of the cosmological constant:
\begin{itemize}

\item
$\Lambda \geq 0$.  The integral of the scalar curvature is strictly positive.
In four dimensions the GB theorem says that
$\oint_{\mathbb{S}^{2}}\tilde{\epsilon}\mathcal{R}=8\pi(1-g)$, with $g$ the genus
of the surface $\mathbb{S}^{2}$.  In this case $\eta<0$ implies that $g=0$ and hence
the only possibility is that the cross sections are two-spheres $S^{2}$.  In five
dimensions $\eta<0$ implies that the cross sections are of positive Yamabe type;
this implies that topologically $\mathbb{S}^{3}$ can only be a finite connected
sum of the three-sphere $S^{3}$ or of the ring $S^{1}\times S^{2}$ (Schoen and Yau
1979; Galloway and Schoen 2006; Galloway 2006).  Both these topologies have been
realized and the corresponding solutions, for example the Myers-Perry black hole
(Myers and Perry 1986) and the Emparan-Reall black ring (Emparan and Reall 2002),
are well known.

\item
$\Lambda < 0$.  The integral of the scalar curvature can have either sign, or even
vanish, and the inequality will always be satisfied.  The only restriction is that
\bea
\oint_{\mathbb{S}^{D-2}}\bm{\tilde{\epsilon}}(\mathcal{R} + 2|\Lambda|)
\geq 2\oint_{\mathbb{S}^{D-2}}\bm{\tilde{\epsilon}}(T_{ab}\ell^{a}n^{b}
+ \|\tilde{\omega}\|^{2}) \; .
\label{adstopologyconstraint}
\eea
There is no constraint on the topology of $\mathbb{S}^{D-2}$ except that the space
has to be compact.  Owing to this special property, many such black holes have been
found with exotic topologies in $D\geq3$ dimensions.  See e.g. (Ba$\tilde{\mbox{n}}$ados
\emph{et al} 1993; $\stackrel{\circ}{\mbox{A}}$minneborg \emph{et al} 1996; Vanzo 1997;
$\stackrel{\circ}{\mbox{A}}$minneborg \emph{et al} 1998; Ba$\tilde{\mbox{n}}$ados 1998;
Ba$\tilde{\mbox{n}}$ados \emph{et al} 1998; Klemm \emph{et al} 1998).

\end{itemize}
For extremal horizons, $\eta=0$, and the constraint (\ref{curvaturecondition})
becomes an equality.  In this case the same restrictions apply to
$\oint_{\mathbb{S}^{D-2}}\bm{\tilde{\epsilon}}\mathcal{R}$ as for nonextremal horizons.
However, there is also a special case that occurs:
\bea
\oint_{\mathbb{S}^{D-2}}\bm{\tilde{\epsilon}}\mathcal{R} = 0
\eea
for an extremal and non-rotating ($\omega=\tilde{\omega}=0$) horizon when the scalar
$T_{ab}\ell^{a}n^{b}$ vanishes on the horizon.  This case corresponds to the torus
topology $T^{D-2}$.

\noindent{\bf Remark.}
Although the expression (\ref{adstopologyconstraint}) does not constrain the topology
of ADS black holes explicitly, there is an interesting area-topology relation that comes
out.  The cosmological term can be integrated out, and upon rearranging to isolate the
surface area
$\mathcal{A}_{\mathbb{S}^{2}} \equiv \oint_{\mathbb{S}^{D-2}}\bm{\tilde{\epsilon}}$ gives
\bea
\mathcal{A}_{\mathbb{S}^{D-2}} \geq \frac{1}{|\Lambda|}\oint_{\mathbb{S}^{D-2}}\bm{\tilde{\epsilon}}
\left(-\frac{1}{2}\mathcal{R} + T_{ab}\ell^{a}n^{b} + \|\tilde{\omega}\|^{2}\right) \; .
\eea
In four dimensions, the GB theorem then implies that
\bea
\mathcal{A}_{\mathbb{S}^{2}}
\geq \frac{1}{|\Lambda|}\left[4\pi(g-1) + \oint_{\mathbb{S}^{2}}\bm{\tilde{\epsilon}}
(T_{ab}\ell^{a}n^{b} + \|\tilde{\omega}\|^{2})\right] \; .
\label{areabound}
\eea
This implies that the maximum allowed angular momentum is bound by the genus and area of the horizon;
see (Booth and Fairhurst 2008; Hennig \emph{et al} 2008) for discussions of the corresponding result
for asymptotically flat spacetimes and appendix B of (Booth and Fairhurst 2008) for a particular
discussion of Kerr-ADS. 

Alternatively, reversing the inequality, one can view it as bounding the allowed area of isolated horizons
from below by the scale of the cosmological curvature and the genus of the horizon: higher genus horizons
necessarily have larger areas. Similar bounds have previously been discovered for stationary ADS black holes
(Gibbons 1999; Woolgar 1999; Cai and Galloway 2001).

\section{Supersymmetric isolated horizons}

\noindent{\emph{``String theorists listening to talks on loop quantum gravity
are often puzzled by the lack of interest in supersymmetry and higher dimensions,
which string theory has shown seem to be required to satisfy certain criteria for
a good theory.''}}
$\sim$ L Smolin

\subsection{Black holes and Killing spinors}

Until now we have discussed the mechanics of  WIHs in arbitrary dimensions.  We now 
specialize to supersymmetric horizons and in particular we focus on the bosonic sector
of four-dimensional $N=2$ gauged supergravity and the bosonic sector of five-dimensional $N=1$
supergravity.  In both cases, black holes are solutions to the bosonic equations of motion
and so the fermion fields vanish. By definition, supersymmetric solutions are invariant under 
the full supersymmetry transformations that are generated by spinor fields.  This means that
for black hole solutions, these transformations should leave the fermion fields unchanged
(and vanishing). Therefore any such black hole solutions must admit a Killing spinor field. 

For full stationary black hole solutions such as those discussed in (Caldarelli and Klemm
1999; Gauntlett \emph{et al} 1999; Gutowski and Reall 2003), the Killing spinor gives rise
to a (timelike) time-translation Killing vector field in the region outside of the black
hole horizon. However, in the quasilocal spirit of the isolated horizon programme we will
only assume the existence of a Killing spinor \emph{on the horizon itself}.  In this case
the spinor will generate a null geodesic vector field that has vanishing twist, shear, and
expansion and this is an allowed $\ell$ on the WIH.

As we did in $\S 2.7$, we will consider fully IHs, which allows for a clear difference
between nonextremal and extremal IHs.  Finally we define a \emph{supersymmetric isolated
horizon} (SIH) as an IH on which the null vector generated by the Killing spinor coincides
(up to a free constant) with the preferred null vector field arising from the IH structure.
As we shall now see these are necessarily extremal as well as having restricted geometry,
rotation, and matter fields. 

\subsection{Killing spinors in four dimensions}

We will first consider the four-dimensional action.  With $D=4$ and $\Lambda=-3/L^{2}$ the
action (\ref{action5}) is the bosonic action of $N=2$ gauged supergravity.  The (extremal)
KN-ADS black hole, which is a solution to the $N=2$ supergravity with the fermion fields
set to zero.  As was shown in (Kosteleck$\acute{\mbox{y}}$ and Perry 1996), the condition
for a supersymmetric KN-ADS black hole in four dimensions to have positive energy is that
\bea
\mathfrak{M} = |\mathfrak{Q}|\left(1 \pm \frac{a}{L}\right)\, ,
\label{bps1}
\eea
which is the extremality condition for the KN-ADS black hole relating the mass $\mathfrak{M}$,
total charge $\mathfrak{Q}\equiv\sqrt{q_{e}^{2}+q_{m}^{2}}$ (with $q_{e}$ and $q_{m}$ the
electric and magnetic charges) and angular momentum $\mathfrak{J}=a\mathfrak{M}$ at infinity.
This is also the saturated Bogomol'ny-Prasad-Sommerfeld (BPS) inequality.  When $\Lambda=0$
the equality
(\ref{bps1}) reduces to (Gibbons and Hull 1982)
\bea
\mathfrak{M}=|\mathfrak{Q}| \, ,
\label{bps2}
\eea
which is the extremality condition for the KN black hole.

For four-dimensional $N=2$ gauged supergravity, we shall employ the conventions of
(Caldarelli and Klemm 2003).  The corresponding (bosonic) action is
\bea
S = \frac{1}{16\pi G_{4}}\int_{\mathcal{M}}\Sigma_{IJ} \wedge \Omega^{IJ}
    + \frac{6}{L^{2}}\bm{\epsilon} - \frac{1}{4}\bm{F} \wedge \star \bm{F} \; .
\label{em}
\eea
The necessary and sufficient condition for supersymmetry with vanishing fermion fields
is that there exists a Killing spinor $\epsilon^{\alpha}$ such that
\bea
\left[\nabla_{\! a} + \frac{i}{4}F_{bc}\gamma^{bc}\gamma_{a}
      +\frac{1}{L}\gamma_{a}\right]\epsilon = 0 \; .
\label{cond1}
\eea
Here, $\gamma^{a}$ are a set of gamma matrices that satisfy the anticommutation rule
\bea
\gamma^{a}\gamma^{b} + \gamma^{b}\gamma^{a} = 2\eta^{ab}
\eea
and the antisymmetry product
\bea
\gamma_{abcd} = \epsilon_{abcd} \; .
\eea
$\gamma_{a_{1}\ldots a_{D}}$ denotes the antisymmetrized product of $D$ gamma matrices.
The spinor $\epsilon$ satisfies the reality condition
\bea
\bar{\epsilon} = i(\epsilon)^{\dagger}\gamma_{0} \, ;
\eea
overbar denotes complex conjugation and $\dagger$ denotes Hermitian conjugation.

From $\epsilon$ one can construct five bosonic bilinears $f$, $g$, $V^{a}$, $W^{a}$ and
$\Psi^{ab}=\Psi^{[ab]}$ where
\bea
f = \bar{\epsilon}\epsilon \, ,
\quad
g = i\bar{\epsilon}\gamma^{5}\epsilon \, ,
\quad
V^{a} = \bar{\epsilon}\gamma^{a}\epsilon \, ,
\quad
W^{a} = i\bar{\epsilon}\gamma^{5}\gamma^{a}\epsilon \, ,
\quad
\Psi^{ab} = \bar{\epsilon}\gamma^{ab}\epsilon \; .
\eea
These are inter-related by several algebraic relations (from the Fierz identities) and differential
equations (from the Killing equation (\ref{cond1})) (Caldarelli and Klemm 2003).  For our purposes the
significant ones are:
\bea
V_{a}V^{a}
&=& -W_{a}W^{a} = -(f^{2} + g^{2}) \, , \label{tnvector1} \\
V^a W_a &=& 0 \label{VdW} \, , \\
g W_{a}
&=& \Psi_{ab} V^b \label{PsiV} \, , \\
f \Psi_{ab} &=& -  \epsilon_{abcd} V^c W^d + \frac{1}{2} g \epsilon_{abcd} \Psi^{cd} \, ,\label{VwW} \\
\nabla_a f
& = & F_{ab} V^b \label{Fl} \, , \\
\nabla_a g
& = & - \frac{1}{L} W_a - \frac{1}{2} \epsilon_{abcd} V^b F^{cd} \label{Wdef} \, ,\\
\nabla_a V_b
&=& \frac{1}{L} \Psi_{ab}
    - f F_{ab} + \frac{g}{2} \epsilon_{abcd} F^{cd} \label{Psi1} \, ,\\
 \nabla_a W_b &=& - \frac{g}{L} g_{a b} - F_{(a}^{\; \; c} \epsilon_{b)cde} \Psi^{de}
  + \frac{1}{4} g_{ab} \epsilon_{cdef} 
  F^{cd} \Psi^{ef}  \, \mbox{and} \label{gradA}
 \\   
\nabla_c \Psi_{ab}
&=& \frac{2}{L} g_{c [a} V_{b]} + 2 F_{[a}^{ \; \; d}\epsilon_{b]dce}W^{e}
    + F_{c}^{\; \; d}\epsilon_{dabe}W^{e} + g_{c[a}\epsilon_{b]def}W^{d}F^{ef} \label{OmegaRel} \; . 
\eea
These are general relations for the existence of a Killing spinor in spacetime.  Although the
Killing spinor may exist in a neighbourhood of the horizon, we only require that it exist
\emph{on the horizon itself}.  Henceforth we specialize by setting $f=g=0$ and at the same time
require that the relations hold on $\Delta$.  Thus, the differential equations
(\ref{Fl})-(\ref{OmegaRel}) are only required to hold when the derivatives are pulled-back
onto the horizon. 

With $f=g=0$,  equation (\ref{tnvector1}) implies that $V^a$ and $W^a$ are both null. On an 
SIH we identify $\ell^a = V^a$ and so condition (\ref{connectionondelta}) together with the differential
constraint (\ref{Psi1}) implies that
\bea
\nabla_{\underleftarrow{a}} \ell_{b} =  \omega_a \ell_b = \frac{1}{L}\Psi_{\underleftarrow{a}b}  \, ,
\label{psipullback}
\eea
and using the skew-symmetry of $\Psi_{ab}$ we can write
\bea
\Psi_{ab} = L (\omega_a \ell_b - \omega_b \ell_a) \, . \label{Psi2}
\eea
Then by equation (\ref{PsiV}) 
 \bea
\ell \lrcorner \omega = 0 \Leftrightarrow \kappa_{(\ell)}=0 \, .\label{kap0}
\eea
Thus, an SIH is necessarily extremal. 

For ease of presentation we now assume that the SIH is foliated into spacelike two-surfaces 
$\mathbb{S}_{v}$. One can always construct such a foliation (and its labelling) so that the associated null 
normal $n\equiv dv$ satisfies $\ell \lrcorner n = -1$ (Ashtekar \emph{et al} 2002). Then the two-metric
on the $\mathbb{S}_{v}$ is given by (\ref{dminustwometric}) and area form on the $\mathbb{S}_{v}$ can be written
as
\bea
\tilde{\epsilon}_{cd} = - \ell^{a}n^{b}\epsilon_{abcd} \; .
\eea
Now we note that with $\kappa_{(\ell)} = 0$ it follows from (\ref{amoneform}) that $\omega_{a}=\tilde{\omega}_{a}$
and hence $\omega_{a} \in T^\star (\mathbb{S}_{v})$.  Finally, with respect to this foliation, the usual restriction
(\ref{pullback1}) and (redundantly) equation (\ref{Fl}) implies that the electromagnetic field takes the form
\bea
F_{ab} = 2E_\perp\ell_{[a}n_{b]} +  B_\perp \tilde{\epsilon}_{ab} + 2\tilde{X}_{[a}\ell_{b]} \, , \label{F}
\label{fdecomp}
\eea
on $\Delta$. Here, $E_\perp$ and $B_\perp$ are the electric and magnetic fluxes through the surface and 
$\tilde{X}^a \in T(\mathbb{S}_{v})$ describes flows of electromagnetic radiation along (but not through) the horizon.

With these preliminaries in hand we can consider the properties of SIHs in asymptotically ADS spacetimes.
First, relations (\ref{VdW}) and (\ref{VwW}) tell us that
\bea
W^a = L \beta V^a
\eea
for some function $\beta$ (the factor of $L$ has been included for later convenience). Then
the pull-back of (\ref{Wdef}) trivially vanishes without giving us any new information but 
(\ref{gradA}) provides a differential equation for $\beta$ on each $\mathbb{S}_{v}$
\bea
d_a \beta + \beta \tilde{\omega}_a = B_\perp \tilde{\omega}_a - E_\perp \tilde{\epsilon}_a^{\; \; b}
 \tilde{\omega}_b \, , 
\eea
where $d_a$ is the intrinsic covariant derivative on $\mathbb{S}_{v}$,
along with its time-invariance: $\pounds_\ell \beta = 0$. 

Next applying the various properties of extremal IHs, one can show that the pull-back of (\ref{OmegaRel})
is 
\bea
\nabla_{\! \underleftarrow{c}} \Psi_{a b} = 2 L \left( \frac{1}{L^2} - \beta B_\perp \right) 
\tilde{q}_{c[a} \ell_{b]} + 2 L \beta E_\perp \tilde{\epsilon}_{c[a} \ell_{b]} \, ,  
\eea
and combining this with (\ref{Psi2}) we find that 
\bea
d_a \tilde{\omega}_b + \tilde{\omega}_a \tilde{\omega}_b
=  \left( \frac{1}{L^2} - \beta B_\perp \right) \tilde{q}_{ab} + \beta E_\perp \tilde{\epsilon}_{ab} \; .
\label{dtom}
\eea

Now as was seen in (\ref{angularmomentum}), the gravitational angular momentum associated with a rotational
Killing field $\phi^a$ is 
\bea
\mathcal{J}_{\rm Grav} = \frac{1}{8 \pi G_4}\oint_{\mathbb{S}_{v}} \bm{\tilde{\epsilon}} \phi \lrcorner \tilde{\omega} \, ,
\eea
and so a necessary condition for non-zero angular momentum is a non-vanishing {rotation one-form} $\tilde{\omega}_a$.
That said, this is not quite sufficient as it is possible for a non-vanishing $\phi \lrcorner \tilde{\omega}$ to
integrate to zero.  For example, consider the case where $\mathbb{S}_{v}$ has topology $S^2$ and $\phi^a$ is a Killing
field (and so divergence-free). Then for some function $\zeta$  we can write $\phi^a = \tilde{\epsilon}^{ab} d_b \zeta$
and
\bea
\oint_{\mathbb{S}_{v}} \bm{\tilde{\epsilon}} \phi \lrcorner \tilde{\omega} = \oint_{\mathbb{S}_{v}}  \zeta d \tilde{\omega} \; .
\eea
Thus, for all closed rotational one-forms ($d \tilde{\omega} = 0$) the associated gravitational angular momentum
will vanish.  As such, it is standard in the isolated horizon literature [see e.g. (Ashtekar \emph{et al} 2001)]
to take $d\tilde{\omega}\neq0$ as the defining characteristic of a rotating isolated horizon. In our case
\bea
 d_{[a} \tilde{\omega}_{b]} =  \beta E_\perp \tilde{\epsilon}_{ab} \; , 
\label{twodcurvature}
\eea
and so an SIH is rotating if and only if $\beta E_\perp \neq 0$. Thus, a rotating horizon must have a
non-trivial electromagnetic field. This is in agreement with known exact solutions: rotating supersymmetric
Kerr-Newmann-AdS black holes as well as those with cylindrical or higher genus horizons all have non-trivial
EM fields (Caldarelli and Klemm 1999).

\subsection{Killing spinors in five dimensions}

We will now consider the five-dimensional action.  With $D=5$ and $\Lambda=0$ the action
(\ref{action5}) is the bosonic sector of $N=1$ gauged supergravity.  As in the four-dimensional EM theory,
solutions to the bosonic equations of motion require the existence of a Killing spinor to ensure that
supersymmetry is preserved.  For black holes, the positive energy theorem together with this requirement
imply that the mass and charge at infinity are constrained such that (Gibbons \emph{el al} 1994)
\bea
\mathfrak{M} = \frac{\sqrt{3}}{2}|\mathfrak{Q}| \; .
\label{bps3}
\eea
As can be verified, the equality (\ref{bps3}) is satisfied by the (5D) extremal RN black hole (Myers and
Perry 1986), the BMPV black hole (Breckenridge \emph{et al} 1997) and the EEMR black ring (Elvang
\emph{et al} 2004).

The strategy for finding supersymmetric solutions to the bosonic equations of motion
based on Killing spinors is essentially the same in five dimensions as it is in four
dimensions.  However, a problem arises specifically in five dimensions -- spinors
satisfying certain reality conditions cannot be consistently defined unless they come
in pairs and are equipped with a symplectic structure.  For details we refer the
interested reader to the excellent Les Houches lectures by van Nieuwenhuizen (1984).

For five-dimensional $N=1$ supergravity, we shall employ the conventions of (Gauntlett
\emph{et al} 2003).  The corresponding (bosonic) action is
\bea
S = \frac{1}{16\pi G_{5}}\int_{\mathcal{M}}\Sigma_{IJ} \wedge \Omega^{IJ}
    - \frac{1}{4}\bm{F} \wedge \star \bm{F} - \frac{2}{3\sqrt{3}}\bm{A} \wedge
    \bm{F} \wedge \bm{F} \; .
\label{emcs}
\eea
The necessary and sufficient condition for supersymmetry with vanishing fermion fields is
that there exists a Killing spinor $\epsilon^{\alpha}$ ($\alpha,\beta,\ldots\in\{1,2\}$)
such that
\bea
%\fl
\left[\nabla_{I} + \frac{1}{4\sqrt{3}}(\Gamma_{I}^{\phantom{a}JK}
- 4\delta_{I}^{\phantom{a}J}\Gamma^{K})F_{JK}\right]\epsilon^{\alpha} = 0 \; .
\label{cond3}
\eea
Here, $\Gamma^{I}$ are a set of gamma matrices that satisfy the anticommutation rule
\bea
\Gamma^{I}\Gamma^{J} + \Gamma^{J}\Gamma^{I} = 2\eta^{IJ}
\eea
and the antisymmetry product
\bea
\Gamma_{IJKLM} = \epsilon_{IJKLM} \; .
\eea
$\Gamma_{I_{1}\ldots I_{D}}$ denotes the antisymmetrized product of $D$ gamma matrices.
The spinors $\epsilon^{\alpha}$ satisfy the reality condition
\bea
\bar{\epsilon}^{\alpha} \equiv (\epsilon^{\alpha})^{\dagger}\Gamma_{0}
= (\epsilon^{\alpha})^{T}\mathcal{C} \, ;
\eea
the second equality is the symplectic Majorana condition, where $T$ denotes
matrix transpose and $\mathcal{C}$ the charge conjugation operator satisfying
\bea
\mathcal{C}(\Gamma_{I})^{T}\mathcal{C}^{-1} = \Gamma_{I} \; .
\eea
Spinor indices are raised and lowered using the symplectic structure
$\epsilon_{\alpha\beta}$ which is defined such that $\epsilon_{12}=\epsilon^{12}=+1$.

Using $\epsilon^{\alpha}$ we can construct bosonic bilinears $\mathcal{F}$, $\mathcal{V}^{I}$
and $\Phi^{\alpha\beta}=\Phi^{(\alpha\beta)}$ such that
\bea
\mathcal{F}\epsilon^{\alpha\beta} = \bar{\epsilon}^{\alpha}\epsilon^{\beta} \, ,
\quad
\mathcal{V}^{I}\epsilon^{\alpha\beta} = \bar{\epsilon}^{\alpha}\Gamma^{I}\epsilon^{\beta} \, ,
\quad
\Phi^{\alpha\beta IJ} = \bar{\epsilon}^{\alpha}\Gamma^{IJ}\epsilon^{\beta} \; .
\eea
As in $\S 3.2$, these bilinears are inter-related by algebraic relations
and differential equations.  For our present purposes we only need the following
(Gauntlett \emph{et al} 2003):
%$\bar{\epsilon}_{1},\epsilon_{2},\bar{\epsilon}_{3}$ and $\epsilon_{4}$,
%given by
%\bea
%\bar{\epsilon}_{1}\epsilon_{2}\bar{\epsilon}_{3}\epsilon_{4}
%= \frac{1}{4}\left(\bar{\epsilon}_{1}\epsilon_{4}\bar{\epsilon}_{3}\epsilon_{2}
%  + \bar{\epsilon}_{1}\Gamma_{I}\epsilon_{4}\bar{\epsilon}_{3}\Gamma^{I}\epsilon_{2}
%  - \frac{1}{2}\bar{\epsilon}_{1}\Gamma_{IJ}\epsilon_{4}\bar{\epsilon}_{3}\Gamma^{IJ}
%  \epsilon_{2}\right) \; .
%\eea
%Of particular interest for our purposes is the identity with
%$\bar{\epsilon}_{1}=\bar{\epsilon}^{\alpha}$, $\epsilon_{2}=\epsilon^{\delta}$,
%$\bar{\epsilon}_{3}=\bar{\epsilon}^{\gamma}$ and $\epsilon_{4}=\epsilon^{\beta}$.
%Then we find that
%\bea
%\epsilon^{\alpha\delta}\epsilon^{\gamma\beta}f^{2}
%= \frac{1}{4}\epsilon^{\alpha\beta}\epsilon^{\gamma\delta}(f^{2} + V_{I}V^{I})
%  - \frac{1}{8}\Phi_{\phantom{aa}IJ}^{\alpha\beta}\Phi^{\gamma\delta IJ} \, ,
%\eea
%and contracting both sides of this equation with
%$\epsilon_{\alpha\beta}\epsilon_{\gamma\delta}$ leads to
\bea
\mathcal{V}_{I}\mathcal{V}^{I} &=& \mathcal{F}^{2} \, , \label{tnvector2}\\
\nabla_{I}\mathcal{V}_{J} &=& \frac{2}{\sqrt{3}}F_{IJ}\mathcal{F} + \frac{1}{2\sqrt{3}}\epsilon_{IJKLM}F^{KL}\mathcal{V}^{M} \; . \label{cond4}
\eea
For IHs, $\mathcal{F}=0$ and $\mathcal{V}^{I}$ is null.  Then, using (\ref{connectionondelta}) together with (\ref{cond4}), and again
making the identification $\mathcal{V}^{a}=\ell^{a}$, we find that an IH of five-dimensional $N=1$ supergravity equipped with a null normal
$\ell$ will be supersymmetric if
\bea
\nabla_{\!\underleftarrow{a}}\ell_{b}
\approx \frac{1}{2\sqrt{3}}e_{\underleftarrow{a}}^{\phantom{a}I}e_{b}^{\phantom{a}J}
        \epsilon_{IJKLM}F^{KL}\ell^{M} \; .
\label{omeganeqzero2}
\eea
It follows that the RHS in (\ref{omeganeqzero2}) vanishes because of the IH condition (\ref{pullback1}) on
$\bm{F}$ and the pullback expression (\ref{pullbackofcoframe}) for $e$, and therefore that $\omega=0$.

\subsection{Interpretation}

In this chapter we examined the restrictions that are imposed on IHs when they are
assumed to be supersymmetric.  The necessary and sufficient condition for supersymmetry
in four-dimensional ADS spacetime is that there exists a Killing spinor $\epsilon$ that
satisfies the conditions (\ref{cond1}).  For four-dimensional SIHs in asymptotically ADS
spacetimes we found that the surface gravity vanishes identically from the algebraic
conditions that are implied by the Killing spinor equation.  This means that the SIHs
are necessarily extremal.  A further constraint that we found for these SIHs is that the
corresponding connection one-form is generically non-zero and is not closed.  Then it
follows from (\ref{twodcurvature}) that these SIHs are rotating for non-trivial
electromagnetic fields.  As we will see below, such a SIH can be non-rotating if the
horizon cross sections are constant curvature surfaces and there is magnetic (but not
electric) charge.

The necessary and sufficient condition for supersymmetry in five dimensions is that there
exists a Killing spinor $\epsilon^{\alpha}$ that satisfies (\ref{cond3}).  For five-dimensional
SIHs in asymptotically flat spacetimes we found that $\omega$ vanishes identically.  This implies
that the corresponding SIHs are non-rotating.  The condition also implies that $\kappa$ is
zero.  The corresponding SIHs are therefore non-rotating and extremal.

These properties impose additional constraints on the topology of the corresponding IHs.
For SIHs in ADS spacetime there is still no constraint on the possible topologies.  The
topologies become severely restricted, however, when $\Lambda=0$.  For SIHs in asymptotically
flat spacetimes the connection $\omega$ vanishes both in four dimensions (see the appendix)
and in five dimensions.  In this case the topology constraint (\ref{curvaturecondition}) gives
\bea
\oint_{\mathbb{S}^{D-2}}\bm{\tilde{\epsilon}}\mathcal{R}
= 2\oint_{\mathbb{S}^{D-2}}\bm{\tilde{\epsilon}}T_{ab}\ell^{a}n^{b} \; .
\eea
In four dimensions we find that there are two possibilities for the topology of a SIH:
\begin{itemize}

\item
If $T_{ab}\ell^{a}n^{b}>0$ then $\oint_{\mathbb{S}^{2}}\bm{\tilde{\epsilon}}\mathcal{R}>0$.
In this case $\mathbb{S}^{2}\cong S^{2}$;

\item
If $T_{ab}\ell^{a}n^{b}=0$ then $\oint_{\mathbb{S}^{2}}\bm{\tilde{\epsilon}}\mathcal{R}=0$.
In this case $\mathbb{S}^{2}\cong T^{2}$.

\end{itemize}
In five dimensions we find that there are three possibilities:
\begin{itemize}

\item
If $T_{ab}\ell^{a}n^{b}>0$ then $\oint_{\mathbb{S}^{3}}\bm{\tilde{\epsilon}}\mathcal{R}>0$.
In this case $\mathbb{S}^{3}\cong S^{3}$ or $\mathbb{S}^{3}\cong S^{1}\times S^{2}$;

\item
If $T_{ab}\ell^{a}n^{b}=0$ then $\oint_{\mathbb{S}^{3}}\bm{\tilde{\epsilon}}\mathcal{R}=0$.
In this case $\mathbb{S}^{3}\cong T^{3}$.

\end{itemize}
Exact solutions to the field equations for the cases where $\mathbb{S}^{2}\cong S^{2}$,
$\mathbb{S}^{3}\cong S^{3}$ and $\mathbb{S}^{3}\cong S^{1}\times S^{2}$ are known.  The
torus topologies, which are classically allowed topologies, have not been found as of yet.

As is the case for spacetimes with no cosmological constant, SIHs in ADS spacetime have vanishing
surface gravity and so are always extremal.  However, in contrast to the asymptotically flat case,
ADS SIHs in four dimensions can be either rotating or non-rotating with strong constraints linking
the rotation to the electromagnetic and Killing spinor fields.  To give a taste of their application,
let us apply these constraints to the case when $\tilde{\omega}=0$.  Then, the Maxwell equations along
with the extremal IH conditions tell us that $E_\perp$ and $B_\perp$ are both constant in time
($\pounds_\ell E_\perp = \pounds_\ell B_\perp = 0$).  In addition, the Maxwell equations
\bea
\nabla_{a}F^{ab} &=& 0\\
\nabla_{[a}F_{bc]} &=& 0
\eea
can be projected to $\mathbb{S}_{v}$, respectively, such that
\bea
\tilde{q}_{a}^{\phantom{a}b}\tilde{q}^{cd}\nabla_{c}F_{db}
- \tilde{q}_{a}^{\phantom{a}b}n^{c}\ell^{d}\nabla_{c}F_{db}
+ \tilde{q}_{a}^{\phantom{a}b}n^{c}\ell^{d}\nabla_{d}F_{bc} &=& 0\\
\tilde{q}_{a}^{\phantom{a}b}n^{c}\ell^{d}\nabla_{b}F_{cd}
+ \tilde{q}_{a}^{\phantom{a}b}n^{c}\ell^{d}\nabla_{c}F_{db}
+ \tilde{q}_{a}^{\phantom{a}b}n^{c}\ell^{d}\nabla_{d}F_{bc} &=& 0 \; .
\eea
Adding these two equations together, and using the decomposition (\ref{fdecomp}) for $F_{ab}$ along
with the assumption $\tilde{\omega}=0$ then gives
\bea
\tilde{q}_{a}^{\phantom{a}b}n^{c}\ell^{d}\nabla_{b}F_{cd}
+ \tilde{q}_{a}^{\phantom{a}b}\tilde{q}^{cd}\nabla_{c}F_{db}
+ 2\tilde{q}_{a}^{\phantom{a}b}n^{c}\ell^{d}\nabla_{d}F_{bc}
= d_a B_\perp + \tilde{\epsilon}_a^{\; \; b} d_b E_\perp = 0 \, . 
\eea
Hence $E_\perp$ and $B_\perp$ are also constant on each $\mathbb{S}_{v}$. Next the supersymmetry constraint
(\ref{dtom}) says that
\bea
\beta B_\perp = \frac{1}{L^2} \; \; \mbox{and} \; \; \beta E_\perp = 0  \, .  
\eea
Thus, $E_\perp = 0$ while $B_\perp \neq 0$ -- that is, these SIHs necessarily have magnetic, but not
electric, charges.  Further, applying the extremality condition (\ref{curvaturecondition}):
\bea
\frac{1}{2} \mathcal{R} &=& d_a \tilde{\omega}^a + \tilde{\omega}_a \tilde{\omega}^a + T_{ab} \ell^a n^b
                            - \frac{3}{L^2} \label{extremality}\\
&=& B^2_\perp - \frac{3}{L^2} \; .
\eea
[This equation has been solved by Kunduri and Lucietti (2008) for vacuum gravity in the context of
near-horizon geometries.]  It is clear that the two-dimensional Ricci curvature $\mathcal{R}$ of the
$\mathbb{S}_{v}$ is constant in this case -- unfortunately the sign of that curvature does not seem to be
determined by the equations.  Consulting a listing of exact supersymmetric black hole solutions (Caldarelli
and Klemm 1999) we see that such solutions are known: specifically there is a supersymmetric asymptotically
ADS black hole in four dimensions which can be non-rotating if the horizon cross sections have genus $g>1$.
As prescribed by our formalism, these solutions have magnetic but not electric charge.

The quasilocal picture that we have presented is in excellent agreement with the results
that are known for stationary spacetimes (Gibbons \emph{et al} 1994; Gauntlett \emph{et al}
1999; Gutowski and Reall 2004).  In that context a supersymmetric black hole in a spacetime
with $\Lambda=0$ (in four and in five dimensions) contains an extremal and non-rotating
horizon.  Extremality is a consequence of the BPS bounds being saturated, which implies
that there exists a Killing spinor.  Non-rotation is then a consequence of the fact that a
vector cannot be constructed in the neighbourhood of a supersymmetric Killing horizon that
is spacelike, as can be seen from the algebraic conditions (\ref{tnvector1}) and
(\ref{tnvector2}).  Therefore the spacetime of such a black hole cannot contain an
ergoregion, which means that the horizon must be non-rotating.  On the other hand, a
supersymmetric black hole with $\Lambda<0$ contains a rotating and extremal horizon (with
non-trivial electromagnetic field).  The rotation is required, otherwise the extremal limit
would result in a naked singularity.

%In this chapter we have established the following generic properties of SIHs in four and in
%five dimensions:

%\noindent{\emph{A SIH of four-dimensional $N=2$ gauged supergravity is extremal, and is
%either: (a) rotating with non-trivial electromagnetic field; or (b) non-rotating with
%constant curvature horizon cross sections and magnetic (but not electric) charge.  A SIH
%of four-dimensional $N=2$ supergravity and of five-dimensional $N=1$ supergravity with
%zero cosmological constant is non-rotating and extremal.}}

In this chapter we focused on null Killing spinors that are defined at the horizon itself.
However, the results obtained here for black holes will not be affected if we assume that
the spinors are defined globally.  We note that there are many other solutions with
such spinors that are defined for the entire spacetime, which do not describe black holes.
These include pp-waves (plane-fronted gravitational waves with parallel rays) -- spacetimes
with vanishing expansion, shear and twist, and are a subset of a wider class of solutions
that share this property, known as Kundt spacetimes.  For details, see e.g. (Stephani
\emph{et al} 2003).  If a spinor is defined globally then $\ell$, which is hypersurface
orthogonal everywhere, is also defined globally.  Therefore such spacetimes can actually
be foliated by SIHs (Pawlowski \emph{et al} 2004).

\subsection{Relation to asymptotically flat solutions}

In five dimensions, there are two supersymmetric solutions with the property that the
bulk electromagnetic field contains angular momentum while the horizon is non-rotating.
These are the BMPV black hole (Breckenridge \emph{et al} 1997) and the EEMR black ring
(Elvang \emph{et al} 2004).

The BMPV black hole was first discovered in (Breckenridge \emph{et al} 1997) as a solution
to the equations that result from the truncation of $D=6$ supergravity.  Later it was shown
that this spacetime is a solution to the $D=5$ EM-CS theory (Gauntlett \emph{et al} 1999).
As was shown in (Reall 2003), the BMPV black hole is the unique asymptotically flat extremal
solution to EM-CS theory with horizon topology $S^{3}$.  The line element of this spacetime
can be written in the following form:
\bea
dS^{2} = -\left(1 + \frac{\mu}{r^{2}}\right)^{-2}\left(dt + \frac{j\sigma_{3}}{2r^{2}}\right)^{2}
         + \left(1 + \frac{\mu}{r^{2}}\right)(dr^{2} + r^{2}d\Omega_{3}^{2}) \, ,
\label{bmpv}
\eea
with
\bea
d\Omega_{3}^{2} &=& \frac{1}{4}(d\theta^{2} + d\phi^{2} + d\psi^{2} + 2\cos\theta d\phi d\psi)\\
\sigma_{3}      &=& d\psi + \cos\theta d\phi \, ;
\eea
the angular coordinates have the ranges
\bea
0\leq \theta < \pi \, ,
\quad
0\leq \phi < 2\pi \, ,
\quad
0\leq \psi < 4\pi \; .
\eea
The ($U(1)$) vector potential that solves the EM-CS equations of motion with the metric
(\ref{bmpv}) is
\bea
\bm{A} &=& \frac{\sqrt{3}}{2}\left[\left(1 + \frac{\mu}{r^{2}}\right)^{-1}
           (dt + \frac{j\sigma_{3}}{2r^{2}}) - dt\right] \; .
\eea
The parameters $\mu$ and $j$ are related to the total mass $\mathfrak{M}$,
charge $\mathfrak{Q}$ and angular momentum $\mathfrak{J}$ at infinity via
\bea
\mathfrak{M} = \frac{3\pi\mu}{4G_{5}} \, ,
\quad
\mathfrak{Q} = \frac{\sqrt{3}\pi\mu}{2G_{5}} \, ,
\quad
\mathfrak{J} = -\frac{\pi j}{2G_{5}} \; .
\eea
Here, the mass is related to the total charge via
\bea
\mathfrak{M} = \frac{\sqrt{3}}{2}\mathfrak{Q} \, ,
\eea
which implies that the BPS bound is saturated.  This is a typical characteristic of
supersymmetric solitons in $D=5$ supergravity (Gibbons \emph{et al} 1994).

The BMPV black hole has one independent rotation parameter.  With
$\mathcal{J}_{\rm Grav}=0$ this corresponds to a SIH with one angular momentum given
by
\bea
\mathcal{J}_{\rm EM} = \frac{1}{8\pi G_{5}}\oint_{S^{3}}
                         (\phi \lrcorner \bm{A})\bm{\Phi}
                     = \frac{j\pi}{4G_{5}}\left(1 - \frac{j^{2}}{\mu^{3}}\right) \; .
\label{bmpvam}
\eea
The spacetime of the BMPV black hole is described by a non-rotating spherical horizon
with angular momentum stored in the electromagnetic fields.  In addition, the distribution
of angular momentum of the spacetime is such that there is a non-zero fraction on the horizon
as well as a negative fraction behind the horizon (Gauntlett \emph{et al} 1999).  The net
result is that the horizon geometry is that of a squashed three-sphere.  Within our
framework, these interesting physical properties correspond to IHs with arbitrary
distortions and rotations in their neighbourhoods.  Such IHs have been studied using
multipole moments (Ashtekar \emph{et al} 2004).  When the angular momentum of the BMPV
black hole is zero the solution reduces to the extremal RN solution in isotropic coordinates.

The generalization of the BMPV black hole to the case where the solution has two
independent angular momentum parameters is the EEMR black ring (Elvang \emph{et al}
2004).  The solution describes an extremal black hole with horizon topology
$S^{1}\times S^{2}$.  The solution was discovered by Elvang \emph{et al} (2004); the
properties and stringy origin were studied in detail in (Elvang \emph{et al} 2005).
The line element of this spacetime can be written in the following form:
\bea
dS^{2} = -f^{2}(dt + \omega_{\psi}d\psi + \omega_{\phi}d\phi)^{2} + f^{-1}dS_{4}^{2} \, ,
\label{eemr}
\eea
with
\bea
dS_{4}^{2} &=& \frac{R^{2}}{(x-y)^{2}}\left[\frac{dx^{2}}{1-x^{2}} + \frac{dy^{2}}{y^{2}-1}
               + (y^{2}-1)d\psi^{2} + (1-x^{2})d\phi^{2}\right]\\
f^{-1} &=& 1 + \frac{Q-q^{2}}{2R^{2}}(x-y) - \frac{q^{2}}{4R^{2}}(x^{2} - y^{2})\\
\omega_{\psi} &=& -\frac{q}{8R^{2}}(1-x^{2})[3Q - q^{2}(3 + x + y)]\\
\omega_{\phi} &=& \frac{3q}{2}(1 + y) + \frac{q}{8R^{2}}(1 - y^{2})[3Q - q^{2}(3 + x + y)] \, ;
\eea
the spatial coordinates have the ranges
\bea
-1\leq x \leq 1 \, ,
\quad
-\infty < y \leq -1 \, ,
\quad
0\leq \psi < 2\pi \, ,
\quad
0\leq \phi < 2\pi \; .
\eea
The parameters $q$ and $Q$ are positive constants that are proportional to the magnetic
dipole moment and total charge, and $R>0$ is the radius of a circle that is parametrized
by $\psi$ at $y=-\infty$.  Also note that the set of coordinates $(x,\phi)$ parametrize
a two-sphere.  Therefore the horizon is ``blown up'' to a ring (topologically
$S^{1}\times S^{2}$) in the $5D$ geometry.  The ($U(1)$) vector potential that solves
the EM-CS equations of motion with the metric (\ref{eemr}) is
\bea
\bm{A} = \frac{\sqrt{3}}{2}\left[f(dt + \omega_{\psi}d\psi + \omega_{\phi}d\phi)
         - \frac{q}{2}[(1 + x)d\phi + (1 + y)d\psi]\right] \; .
\eea
The parameters $q$ and $Q$ are related to the total mass $\mathfrak{M}$ and total
charge $\mathfrak{Q}$ at infinity via
\bea
\mathfrak{M} = \frac{3\pi Q}{4G_{5}} \, ,
\quad
\mathfrak{Q} = \frac{\sqrt{3}\pi Q}{2G_{5}} \, ,
\eea
and to the angular momenta $\mathfrak{J}_{\psi}$ and $\mathfrak{J}_{\phi}$ at infinity
via
\bea
\mathfrak{J}_{\psi} = \frac{\pi q}{8G_{5}}(6R^{2} + 3Q - q^{2}) \, ,
\quad
\mathfrak{J}_{\phi} = \frac{\pi q}{8G_{5}}(3Q - q^{2}) \; .
\eea
Note that $\mathfrak{M}=(\sqrt{3}/2)\mathfrak{Q}$ as for the BMPV black hole.  Also
note that when $R=0$ the angular momenta $\mathfrak{J}_{\psi}$ and $\mathfrak{J}_{\phi}$
are equal.  This would suggest that the BMPV black hole with equal angular momenta is
essentially the EEMR black ring in the limit when $R=0$.  However, in this limit there
is then an apparent singularity in the four-metric and also in the connection one-form.

The EEMR black ring has two independent rotation parameters.  This corresponds to a SIH
with two angular momenta given by
\bea
\mathcal{J}_{j} = \frac{1}{8\pi G_{5}}\oint_{S^{1}\times S^{2}}
                  (\phi_{j} \lrcorner \bm{A})\bm{\Phi}
\quad
(j\in\{1,2\}) \; .
\eea
In addition, a black ring can also have dipole charges which are naturally defined
on the horizon (Astefanesei and Radu 2006; Copsey and Horowitz 2006; Emparan and
Reall 2006).  For an IH with ring topology a definition for the dipole charge
$\mathcal{P}$ can be realized by integrating the electromagnetic field strength
plus the CS contribution over $S^{2}$:
\bea
\mathcal{P} = \frac{1}{2\pi}\int_{S^{2}} \star \bm{\Phi} \; .
\eea
Charges of this type appear in the first law for a black ring (Astefanesei and Radu
2006; Copsey and Horowitz 2006).  However, this is not the case with the first law
(\ref{firstlaw}) that we derived.  This is probably due to the fact that the dipole
charges $\mathcal{P}$ are associated with the presence of \emph{magnetic charge},
which we have not incorporated into our current phase space.

\subsection*{Appendix: an alternate formalism in four dimensions}

In four dimensions, there is an alternative formalism that can be used to derive
the supersymmetry conditions for IHs as was originally done for flat spacetime
(Liko and Booth 2008).  This is a NP-type spinor formalism, in which the necessary
and sufficient condition for supersymmetry with vanishing fermion fields is that
there exists a Killing spinor
$\psi_{AA^{\prime}}=(\alpha_{A},\beta_{A^{\prime}})$
($A,B,\ldots\in\{1,2\}$ and $A^{\prime},B^{\prime},\ldots\in\{1,2\}$) such that
(Tod 1983; Tod 1995)
\bea
\nabla_{AA^{\prime}}\alpha_{B} + \sqrt{2}\phi_{AB}\beta_{A^{\prime}} &=& 0
\label{cond5}\\
\nabla_{AA^{\prime}}\beta_{B^{\prime}} - \sqrt{2}\bar{\phi}_{A^{\prime}B^{\prime}}
\alpha_{A} &=& 0 \; .
\label{cond6}
\eea
Here, $\phi_{AB}$ is the Maxwell spinor and $\bar{\phi}_{A^{\prime}B^{\prime}}$ is its
complex conjugate.  These are related to the field strength via
\bea
\bm{F} = \phi_{AB}\epsilon_{A^{\prime}B^{\prime}}
         + \bar{\phi}_{A^{\prime}B^{\prime}}\epsilon_{AB} \, ;
\eea
the spinor symplectic structure is defined such that $\epsilon^{12}=-\epsilon^{21}=1$.
Using the spinors $\alpha$ and $\beta$ we can define the following set of null vectors:
\bea
\ell_{AA^{\prime}} = \alpha_{A}\bar{\alpha}_{A^{\prime}} \, ,
\quad
n_{AA^{\prime}} = \bar{\beta}_{A}\beta_{A^{\prime}} \, ,
\quad
\vartheta_{AA^{\prime}} = \alpha_{A}\beta_{A^{\prime}} \, ,
\quad
\bar{\vartheta}_{AA^{\prime}} = \bar{\beta}_{A}\bar{\alpha}_{A^{\prime}} \; .
\label{spinvectors}
\eea
It can be shown that the vector
\bea
K_{AA^{\prime}} \equiv \ell_{AA^{\prime}} + n_{AA^{\prime}}
\eea
is a Killing vector; the norm of this vector is given by
\bea
\| K\| = 2V\bar{V} \, ,
\label{tnvector3}
\eea
where we defined the scalar $V=\alpha_{A}\bar{\beta}^{A}$.  It follows that
$K_{AA^{\prime}}$ can be either timelike (referred to as nondegenerate) when
$V\neq0$ or null (referred to as degenerate) when $V=0$.

For IHs, the case of interest is the one for which the Killing spinor is null.
This is a particularly special case because $V=\alpha_{A}\bar{\beta}^{A}=0$
implies that
\bea
\bar{\beta}^{A} = \mathcal{K}\alpha^{A}
\eea
for some function $\mathcal{K}$.  Putting this into the conditions (\ref{cond5})
and (\ref{cond6}) allows one to find an expression for the covariant derivative
in terms of the spinors (Tod 1983; Tod 1995):
\bea
\nabla_{AA^{\prime}}\alpha_{B}
= -\sqrt{2}\bar{\mathcal{K}}\phi\alpha_{A}\alpha_{B}\bar{\alpha}_{A^{\prime}} \; .
\eea
Here, $\phi$ is a function defined by $\phi_{AB}=\phi\alpha_{A}\alpha_{B}$.  Let
us use this form of the covariant derivative to find the covariant derivative of the
null normal $\ell$ of an IH.  We find that
\bea
\nabla_{a}\ell_{b} = -\sqrt{2}(\bar{\mathcal{K}}\phi + \mathcal{K}\bar{\phi})
                     \ell_{a}\ell_{b} \; .
\eea
This immediately implies that
\bea
\nabla_{\!\underleftarrow{a}}\ell_{b}\approx0 \, ,
\eea
and with (\ref{connectionondelta}) it follows that $\omega=0$.

\section{Isolated Horizons in EGB Theory}

\noindent{\emph{``Higher-derivative theories are frequently avoided because of
undesirable properties, yet they occur naturally as corrections to general
relativity and cosmic strings.''}}
$\sim$ J Z Simon

\subsection{First-order action for EGB theory}

As in Section 2, we work in the first-order connection-dynamics formulation.  Here,
the configuration space consists of the pair $(e^{I},A_{\phantom{a}J}^{I})$ (with
electromagnetic fields zero).  In this configuration space, the action (\ref{action2})
for EGB theory on the manifold $(\mathcal{M},g_{ab})$ (assumed for the moment to have
no boundaries) is given by
\bea
S = \frac{1}{2\kappa_{D}}\int_{\mathcal{M}}\Sigma_{IJ} \wedge \Omega^{IJ}
    - 2\Lambda\bm{\epsilon}
    + \alpha\Sigma_{IJKL} \wedge \Omega^{IJ} \wedge \Omega^{KL} \; .
\label{action4}
\eea
The equation of motion for the connection is
\bea
\mathscr{D}(\Sigma_{IJ} + 2\alpha\Sigma_{IJKL} \wedge \Omega^{KL}) = 0 \; .
\label{eom4}
\eea
This equation says that, in general, there exists a non-vanishing torsion
$T^{I}=\mathscr{D}e^{I}$.  To see what constraints are imposed on $T$, we
can use the Bianchi identity $\mathscr{D}\Omega^{IJ}=0$ together with the
identity
\bea
\mathscr{D}\Sigma_{I_{1} \ldots I_{m}}
= \mathscr{D}e^{M} \wedge \Sigma_{I_{1} \ldots I_{m}M} \; .
\eea
Substituting these into equation (\ref{eom2}) gives
\bea
T^{I} \wedge (\Sigma_{IJK} + 2\alpha\Sigma_{IJKLM} \wedge \Omega^{LM}) = 0 \; .
\label{constraint}
\eea
In analogy with Einstein gravity, we assume directly that the torsion in
(\ref{constraint}) vanishes.  (The torsion in Einstein gravity is zero, but
this is not an assumption.  The condition follows directly from the equation
of motion for the connection.)  To get the equation of motion for the co-frame
we note that the variation of $\Sigma$ is given by
\bea
\delta\Sigma_{I_{1} \ldots I_{m}}
= \delta e^{M} \wedge \Sigma_{I_{1} \ldots I_{m}M} \; .
\eea
This leads to
\bea
\Sigma_{IJK} \wedge \Omega^{JK} - 2\Lambda\Sigma_{I}
+ \alpha\Sigma_{IJKLM} \wedge \Omega^{JK} \wedge \Omega^{LM} = 0 \; .
\label{eom5}
\eea
The equations (\ref{eom4}) and (\ref{eom5}) for the connection and co-frame are
equivalent to the field equations (\ref{egbfieldequations}) in the metric formulation.

\subsection{Modified boundary conditions and the zeroth law}

Let us now turn to the case for which the manifold $(\mathcal{M},g_{ab})$ has
boundaries; the region of spacetime that we consider for EGB theory is essentially
the same as that which we considered in Section 2 for EM-CS theory.  Namely,
$(\mathcal{M},g_{ab})$ is a $D$-dimensional Lorentzian manifold with topology
$R\times M$, contains a $(D-1)$-dimensional null surface $\Delta$ as inner boundary
(representing the horizon), and is bounded by $(D-1)$-dimensional manifolds $M^{\pm}$
that extend from $\Delta$ to infinity.  As in Section 3, we consider the purely
quasilocal case and neglect any subleties that are associated with the outer boundary.

Let us now state the modification that is required for EGB theory.  First, we note
that for IHs in general relativity the dominant energy condition can be imposed
interchangably onto the Ricci tensor or the stress-energy tensor.  This is because
the Einstein field equations $G_{ab}=2T_{ab}$ imply that
$R_{ab}v^{a}v^{b}=2T_{ab}v^{a}v^{b}$ for any null vector $v^{a}$.  For IHs in EGB
theory this is no longer true because $G_{ab}\neq \kappa_{D}T_{ab}$.  [This is also the
reason why the topology constraint (\ref{constraint}) is not applicable to IHs in EGB
theory.]  It turns out that condition (c) of Definition I for IHs in EGB theory has to
be imposed onto the Ricci tensor in order for the shear tensor to vanish when the
Raychaudhuri equation is employed.  Thus we have the following modified definition for
a WIH in EGB theory:

\noindent{\bf Definition III.}
\emph{A WIH $\Delta$ in EGB theory is a null surface and has a degenerate metric $q_{ab}$
with signature $0+\ldots+$ (with $D-2$ nondegenerate spatial directions) along with an
equivalence class of null normals $[\ell]$ (defined by $\ell \sim \ell^\prime \Leftrightarrow
\ell' = z \ell$ for some constant $z$) such that the following conditions hold: (a) the
expansion $\theta_{(\ell)}$ of $\ell_{a}$ vanishes on $\Delta$; (b) the field equations hold
on $\Delta$; (c) the Ricci tensor is such that the vector $-R_{\phantom{a}b}^{a}\ell^{b}$ is
a future-directed and causal vector; (d) $\pounds_{\ell}\omega_{a}=0$ and
$\pounds_{\ell}\underleftarrow{\bm{A}}=0$ for all $\ell\in[\ell]$.}

The condition (c) on the Ricci tensor is the only modification that needs to be made for WIHs
in EGB theory.  In particular the zeroth law now follows just as it does for IHs in EM-CS theory.
Remarkably, the zeroth law is essentially independent of the functional content of the action,
and follows from the boundary conditions alone.  This is the same conclusion that was obtained
for globally stationary spacetimes (Wald 1993; Iyer and Wald 1994; Jacobson \emph{et al} 1994).

\subsection{Variation of the boundary term}

We have seen that the definition for a nonexpanding horizon needs to be modified
for EGB gravity by imposing the analogue of the dominant energy condition directly
on the Ricci tensor.  In the action principle, the main modification to the formalism
is the appearance of an additional surface term.  Let us therefore reconsider the
action (\ref{action4}) but for a region of the manifold $\mathcal{M}$ that is bounded
by null surface $\Delta$ and spacelike surfaces $M^{\pm}$ which extend to the (arbitrary)
boundary $\mathscr{B}$.

Denoting the pair $(e,A)$ collectively as a generic field variable $\Psi$, the
first variation gives
\bea
\delta S = \frac{1}{2\kappa_{D}}\int_{\mathcal{M}}E[\Psi]\delta\Psi
           + \frac{(-1)^{D}}{2\kappa_{D}}\int_{\partial\mathcal{M}}J[\Psi,\delta\Psi] \; .
\label{firstprime}
\eea
Here $E[\Psi]=0$ symbolically denotes the equations of motion and
\bea
J[\Psi,\delta\Psi] = \widetilde{\Sigma}_{IJ} \wedge \delta A^{IJ}
\label{surfaceprime}
\eea
is the surface term, with $(D-2)$-form
\bea
\widetilde{\Sigma}_{IJ} = \Sigma_{IJ} + 2\alpha\Sigma_{IJKL} \wedge \Omega^{KL} \; .
\label{sigmawidetildeprime}
\eea
If the integral of $J$ on the boundary $\partial\mathcal{M}$ vanishes then the
action principle is said to be differentiable.  We must show that this is the
case.  Because the fields are held fixed at $M^{\pm}$ and at $\mathscr{B}$, $J$
vanishes there.  So we only need to show that $J$ vanishes at the inner boundary
$\Delta$.  To show that this is true we need to find an expression for $J$ in
terms of $A$ and $\widetilde{\Sigma}$ pulled back to $\Delta$.  As for EM-CS theory
we do this by fixing an internal basis consisting of the (null) pair $(\ell,n)$
and $D-2$ spacelike vectors $\vartheta_{(i)}$ given by (\ref{npbasis}) together
with the conditions (\ref{npconditions}).

The pull-back of $A$ is the same as for EM-CS theory, and is given by
(\ref{pullbackofa}).  For EGB theory we also need the pull-back of the curvature
$\Omega$ to $\Delta$, which can be obtained either by direct calculation from
(\ref{pullbackofa}) or by construction from the definition of the Riemann tensor.
We will take the second approach here.  We will use the definition
(\ref{dminustwometric}) of the metric on $\mathbb{S}^{D-2}$ and the definition
(\ref{amoneform}) of the connection projected onto $\mathbb{S}^{D-2}$.

In thinking about these quantities it is useful to keep in mind the case where
$\Delta$ is foliated into spacelike $(D-2)$-surfaces $\mathbb{S}_{v}$ which are labelled
by a parameter $v$ and $n$ is chosen to be $-dv$.  Then $\ell^{a}$ evolves the
foliation surfaces while $\ell^{a}$ and $n^{a}$ together span the normal bundle
$T^{\perp}(\mathbb{S}_{v})$ on which $\tomega_{a}$ is the connection.  Furthermore, the
$\vartheta_{(i)}^{a}$ span the tangent bundle $T(\mathbb{S}_{v})$ and $\tq_{ab}$ is the
metric tensor for the $\mathbb{S}_{v}$.  Then, it is clear that $\Delta$ is non-rotating
when $\tilde{\omega}=0$ provided that the rotational vector field $\phi$ is tangential
to $\Delta$.

We now turn to the Riemann tensor with the first two indices pulled back to $\Delta$. 
By definition,
\bea
R_{\underleftarrow{ab}\phantom{a}d}^{\phantom{aa}c}\ell^{d}
= -q_{a}^{\phantom{a}e}q_{b}^{\phantom{a}f}
  (\nabla_{e}\nabla_{f} - \nabla_{f}\nabla_{e})\ell^{c} \, , 
\label{defriemann}
\eea
and with the horizon identity $\nabla_{\underleftarrow{a}}\ell^{b}=\omega_{a}\ell^{b}$
along with the decomposition (\ref{amoneform}), a few lines of algebra gives
\bea
R_{\underleftarrow{ab}\phantom{a}d}^{\phantom{aa}c}\ell^{d} 
= \left(-2n_{[a} d_{b]}\kappa_{(\ell)} + 2\tq_{[a}^{\phantom{a}e}\tq_{b]}^{\phantom{a}f}
  d_{e}\tomega_{f}
  - 2n_{[a}\tq_{b]}^{\phantom{a}f}\mathcal{L}_{\ell}\tomega_{f}\right)\ell^{c} \, , 
\eea
where $d_{a}$ is the covariant derivative that is compatible with the metric $\tq_{ab}$.
For a weakly isolated horizon the zeroth law ensures that $d_{a}\kappa_{(\ell)}=0$ and
if the horizon is non-rotating then $\tomega=0$ also, whence
\bea
R_{\underleftarrow{ab}\phantom{a}d}^{\phantom{aa}c}\ell^{d} = 0 \; . 
\label{Rl}
\eea
Finally, using this result and (\ref{dminustwometric}) it is straightforward to see that 
\bea
R_{\underleftarrow{ab}cd} \vartheta^c_{(i)} \vartheta^d_{(j)} 
= \tq_{a}^{\phantom{a}e}\tq_{b}^{\phantom{a}f}R_{efcd}\vartheta^c_{(i)}\vartheta^d_{(j)} \;  . 
\eea
From here one can use the fact that 
\bea
d_{a}d_{b}\vartheta^{(i)}_{c}
= \tq_{a}^{\phantom{a}d}\tq_{b}^{\phantom{a}e} \tq_{c}^{\phantom{a}f}
  \nabla_{d}\left(\tq_{e}^{\phantom{a}g}\tq_{f}^{\phantom{a}h}\nabla_{g}\vartheta^{(i)}_{h}\right) \, , 
\eea
and the identity for the Riemann tensor $\mathcal{R}_{abcd}$ associated with 
$\tq_{ab}$ 
\bea
\mathcal{R}_{abcd} \vartheta^{(i)d}
= \left(d_{a}d_{b} - d_{b}d_{a}\right)\vartheta_c^{(i)} \, , 
\eea
along with (\ref{dminustwometric}) to show the Gauss relation
\bea
\tq_{a}^{\phantom{a}e}\tq_{b}^{\phantom{a}f}\tq_{c}^{\phantom{a}g}\tq_{d}^{\phantom{a}h}R_{efgh} 
= \mathcal{R}_{abcd} + \left(k^{(\ell)}_{ac}k^{(n)}_{bd} + k^{(n)}_{ac}k^{(\ell)}_{bd})
-  (k^{(\ell)}_{bc}k^{(n)}_{ad} + k^{(n)}_{bc}  k^{(\ell)}_{ad}\right) \; .
\label{Gauss}
\eea
Here $k^{(\ell)}_{ab}=\tq_{a}^{\phantom{a}c}\tq_{b}^{\phantom{a}d}\nabla_{c}\ell_{d}$
and $k^{(n)}_{ab}=\tq_{a}^{\phantom{a}c}\tq_{b}^{\phantom{a}d}\nabla_{c}n_{d}$ are the
extrinsic curvatures associated with $\ell_{a}$ and $n_{a}$.  However,
$k^{(\ell)}_{ab}=(1/2)\vartheta_{(\ell)}\tq_{ab}+\sigma_{ab}$, and on a non-expanding
horizon both the expansion and shear vanish. Thus for the cases in which we are
interested
\bea
\tq_{a}^{\phantom{a}e}\tq_{b}^{\phantom{a}f}\tq_{c}^{\phantom{a}g}\tq_{d}^{\phantom{a}h}R_{efgh} 
= \mathcal{R}_{abcd} \; .
\label{GaussNEH}
 \eea

Then upon expanding the frame indices of $\Omega_{\underleftarrow{ab}}^{\phantom{aa}IJ}$ in
terms of the $\ell^{I}$, $n^{I}$ and $\vartheta_{(i)}^{\phantom{a}I}$, and applying (\ref{Rl})
and (\ref{GaussNEH}), it follows that on any non-rotating WIH the pull-back of the associated
curvature is
\bea
\Omega_{\underleftarrow{ab}}^{\phantom{aa}IJ}
\approx \vartheta^{(k)}_{a}\vartheta^{(l)}_{b}\mathcal{R}_{kl}^{\phantom{aa}ij}
        \vartheta_{(i)}^{\phantom{a}I}\vartheta_{(j)}^{\phantom{a}J}
        + 2\ell^{[I}\vartheta_{(i)}^{\phantom{a}J]}
          \Omega_{\underleftarrow{ab}}^{\phantom{aa}KL}
          \vartheta_{\phantom{a}K}^{(i)} n_L \, ,
\label{curvature}
\eea
where $\mathcal{R}_{kl}^{\phantom{aa}ij}$ is the Riemann tensor associated
with the $(D-2)$ metric $\tilde{q}_{ab} = g_{ab} + \ell_a n_b + n_a \ell_b$. 
That is, given a foliation of $\Delta$ into spacelike $(D-2)$-surfaces, the
spacelike $\vartheta_{(i)}^{a}$ give an orthonormal basis on those surfaces and 
$\mathcal{R}_{kl}^{\phantom{aa}ij} $ is the corresponding curvature tensor; 
for a non-expanding horizon, these quantities are independent of both the 
slice of the foliation and the particular foliation itself. 

To find the pull-back to $\Delta$ of $\widetilde{\Sigma}$, we use the
decomposition (\ref{pullbackofcoframe}), whence the $(D-2)$-form given by
(\ref{pullbackofsigmaij}) and in $D\geq5$ dimensions, the $(D-4)$-form
\bea
\underleftarrow{\Sigma}_{IJKL}
&\approx&
-\frac{1}{(D-5)!} 
\epsilon_{IJKLA_1 \dots A_{D-4}} \ell^{A_1} 
\vartheta^{\phantom{a}A_2}_{(i_1)} \dots \vartheta^{\phantom{a}A_{D-4}}_{(i_{D-5})}
\left[n \wedge \vartheta^{(i_1)} \wedge \dots \wedge \vartheta^{(i_{D-5})}\right]\nonumber\\
& &
+\frac{1}{(D-4)!} 
\epsilon_{IJKLA_1 \dots A_{D-4}}  
\vartheta^{\phantom{a}A_1}_{(i_1)} \dots \vartheta^{\phantom{a}A_{D-4}}_{(i_{D-4})}
\left[\vartheta^{(i_1)} \wedge \dots \wedge \vartheta^{(i_{D-4})} \right] \; .\nonumber\\
\label{pullbackofsigmaijkl}
\eea
In four dimensions $\underleftarrow{\Sigma}_{IJKL} = \epsilon_{IJKL}$.

These expressions are somewhat formidable but on combining them to find
$\widetilde{\Sigma}_{IJ} \wedge \delta A^{IJ}$ there is significant simplification. 
The key is to note that each term includes a total contraction of
$\epsilon_{I_1 \dots I_{D}}$. This contraction must 
include one copy of each of $\ell^I$, $n^I$, and the $\vartheta^{\phantom{a}I}_{(i)}$
-- else that term will be zero. Similarly the resulting $(D-1)$ form must be proportional
to $n \wedge \vartheta^{(2)} \wedge \dots \wedge \vartheta^{(D-1)}$. Then (\ref{surfaceprime})
becomes
\bea
J[\Psi,\delta\Psi]
\approx \bm{\tilde{\epsilon}} \wedge \delta \omega 
& + & \frac{2\alpha}{(D-4)!}\left[\epsilon_{IJKLA_1\dots A_{D-4}} \ell^I n^J 
\vartheta^{\phantom{a}K}_{(k)} \vartheta^{\phantom{a}L}_{(l)} 
\vartheta^{\phantom{a}A_1}_{(i_1)} \dots \vartheta^{\phantom{a}A_{D-4}}_{i_{D-4}}\right]\nonumber\\
& &
\times \mathcal{R}_{mn}^{\phantom{mn}kl} \vartheta^{(i_1)}
       \wedge \dots \wedge \vartheta^{(i_{D-4})} \wedge \vartheta^{(m)} \vartheta^{(n)}
       \wedge \delta \omega \; .\nonumber\\
\eea
The first and second terms respectively come from the
$\underleftarrow{\Sigma}_{IJ}$ and $\underleftarrow{\Sigma}_{IJKL}$ parts of
$\underleftarrow{\widetilde{\Sigma}}_{IJ}$, $\bm{\tilde{\epsilon}}$ is the area
element defined by (\ref{areaelement}), and we keep in mind that the horizon is
non-rotating so that $\omega_{a}=-\kappa_{(\ell)}n_{a}$.  The second term therefore
also simplifies.  Given that there are only $(D-4)$ elements in the spacelike basis
it is reasonably easy to see that this term sums over cases where $(m,n)$ and $(i,j)$
are the same set of indices. That is (up to a numerical factor) the second term
amounts to contracting $m$ with $i$ and $n$ with $j$ so that the full surface term
reduces to
\bea
J[\Psi,\delta\Psi]
\approx \bm{\tilde{\epsilon}}(1 + 2\alpha\mathcal{R}) \wedge \delta \omega \; . 
\label{simplifiedpullbackofcurrentprime}
\eea
It follows that $J\approx0$ because $\delta\omega=0$ on the initial and final
cross sections of $\Delta$ (i.e. on $M^{-}\cap\Delta$ and on $M^{+}\cap\Delta$),
and because $\delta\omega$ is Lie dragged on $\Delta$.  Therefore the surface term
$J|_{\partial\mathcal{M}}=0$ for EGB gravity, and we conclude that the equations of
motion $E[\Psi]=0$ follow from the action principle $\delta S=0$.

\subsection{Covariant phase space and the first law}

In order to derive the first law we need to find the symplectic structure on
the covariant phase space $\bm{\Gamma}$ consisting of solutions $(e,A)$ to the
EGB field equations on $\mathcal{M}$.  We find that the second variation of the
EGB surface term (\ref{surfaceprime}) gives
\bea
J[\Psi,\delta_{1}\Psi,\delta_{2}\Psi]
= (-1)^{D}\left[\delta_{1}\widetilde{\Sigma}_{IJ} \wedge \delta_{2}A^{IJ}
                - \delta_{2}\widetilde{\Sigma}_{IJ} \wedge \delta_{1}A^{IJ}\right] \, ;
\eea
integrating over $M$ defines the \emph{bulk} symplectic structure
\bea
\bm{\Omega}_{\rm B}(\delta_{1},\delta_{2})
= \frac{(-1)^{D}}{2\kappa_{D}}\int_{M}
  \left[\delta_{1}\widetilde{\Sigma}_{IJ} \wedge \delta_{2}A^{IJ}
   - \delta_{2}\widetilde{\Sigma}_{IJ} \wedge \delta_{1}A^{IJ}\right] \; .
\label{bulksymplecticprime}
\eea
We also need to find the pull-back of $J$ to $\Delta$ and add the
integral of this term to $\bm{\Omega}_{\rm B}$ to determine the full
symplectic structure.  From (\ref{simplifiedpullbackofcurrentprime}) we
have
\bea
\bm{\Omega}_{\rm S}
\approx \frac{(-1)^{D}}{\kappa_{D}}\int_{\Delta}
\left[\delta_{1}\left[\bm{\tilde{\epsilon}}(1+2\alpha\mathcal{R})\right]
\wedge \delta_{2}\omega
- \delta_{2}\left[\bm{\tilde{\epsilon}}(1+2\alpha\mathcal{R})\right]
\wedge \delta_{1}\omega\right] \; .
\eea
which is a total derivative.  Now, with the definition (\ref{potentials}) for $\psi$,
the surface symplectic structure $\Omega_{\rm S}$ is a total derivative, and hence
upon using the Stokes theorem, becomes an integral over $\mathbb{S}^{D-2}$.  The full
symplectic structure for EGB gravity is therefore
\bea
\bm{\Omega}(\delta_{1},\delta_{2})
&=& \frac{1}{2\kappa_{D}}\int_{M}
    \left[\delta_{1}\widetilde{\Sigma}_{IJ} \wedge \delta_{2}A^{IJ}
    - \delta_{2}\widetilde{\Sigma}_{IJ} \wedge \delta_{1}A^{IJ}\right]\nonumber\\
& & + \frac{1}{\kappa_{D}}\oint_{\mathbb{S}^{D-2}}
    \left[\delta_{1}\left[\bm{\tilde{\epsilon}}(1+2\alpha\mathcal{R})\right]
    \wedge \delta_{2}\psi
    - \delta_{2}\left[\bm{\tilde{\epsilon}}(1+2\alpha\mathcal{R})\right]
    \wedge \delta_{1}\psi\right] \, ,
\label{fullsymplecticprime}
\eea
where we have absorbed the overall (irrelevant) factor of $(-1)^{D}$.

We can now proceed to derive the first law.  As before, this follows upon evaluating
the symplectic structure at $(\delta,\delta_{\xi})$, giving a surface term at infinity
(which is identified with the ADM energy) and a surface term at the horizon.  We find
that the surface term at the horizon is given by
\bea
\bm{\Omega}|_{\Delta}(\delta,\delta_{t})
= \frac{\kappa_{(z\ell)}}{\kappa_{D}}\delta\oint_{\mathbb{S}^{D-2}}
  \bm{\tilde{\epsilon}}(1+2\alpha\mathcal{R}) \; .
\eea
where we used $\kappa_{(z\ell)}=\pounds_{z\ell}\psi=z\ell\lrcorner\omega$.  Finally,
assuming that this is a total variation, i.e. that there exists a function $\mathcal{E}$
such that $\bm{\Omega}|_{\Delta}(\delta,\delta_{\xi})=\delta\mathcal{E}$, we find that
\bea
\delta\mathcal{E} = \frac{\kappa_{(z\ell)}}{\kappa_{D}}\delta\oint_{\mathbb{S}^{D-2}}\bm{\tilde{\epsilon}}
                    (1 + 2\alpha\mathcal{R}) \; .
\label{firstlaw5}
\eea
This is the first law for a WIH in EGB theory.  Comparing this to the standard first law
identifies the entropy as
\bea
\mathcal{S} = \frac{1}{4G_{D}}\oint_{\mathbb{S}^{D-2}}\bm{\tilde{\epsilon}}(1 + 2\alpha\mathcal{R}) \; .
\label{entropy4}
\eea
Therefore WIHs in $D$-dimensional EGB theory satisfy the first law (and the zeroth law)
of black-hole mechanics.

\subsection{Comparison with Euclidean and Noether charge methods}

The quasilocal expression for the entropy is in exact agreement with the Noether
charge expression that was derived by Clunan \emph{et al} (2004).  As in that
approach, no assumptions about the cross sections $\mathbb{S}^{D-2}$ of the horizon need
to be made.  An important difference, however, is that we did not assume the
existence of a globally-defined Killing vector.  Instead we had to specify the
existence of a time translation vector field which mimics the properties of a
Killing vector but \emph{is not defined for the entire spacetime}.

In order to compare the entropy to the Euclidean method, we need to reference a
black hole solution.  The EGB equations admit the following class of (static) black
hole solutions (Cai 2002; Cho and Neupane 2002):
\bea
dS^{2} &=& -h(r)dt^{2} + \frac{dr^{2}}{h(r)} + r^{2}d\Omega_{(k)D-2}^{2}\nonumber\\
h(r) &=& k + \frac{r^{2}}{2\tilde{\alpha}}\left(1 - \sqrt{1
             - \frac{8\tilde{\alpha}\Lambda}{(D-1)(D-2)}
             + \frac{8\kappa_{D}\tilde{\alpha}M}{(D-2)\mathscr{V}_{(k)D-2}r^{D-1}}}\right) \; .\nonumber\\
\label{bhsolution}
\eea
Here, $\mathscr{V}_{(k)N-1}=\pi^{N/2}/\Gamma(N/2+1)$ is the volume of an
$(N-1)$-dimensional space $S^{N-1}=S_{(k)}^{N-1}$ of constant curvature with
metric $d\Omega_{(k)N-1}^{2}$; $k$ is the curvature index with $k=1$ corresponding to
positive constant curvature, $k=-1$ corresponding to negative constant curvature, and
$k=0$ corresponding to zero curvature.  $M$ is the mass of the black hole, and
$\tilde{\alpha}$ is related to the GB parameter via
\bea
\tilde{\alpha} = (D-3)(D-4)\alpha \; .
\eea
The singular surfaces with radii $r_{*}$ are given by the roots to the
equation $h(r=r_{*})=0$.  We denote the event horizon by $r_{+}$.  The
location of this surface depends on the sign of the cosmological constant:
if $\Lambda\leq0$ then the largest root $r_{+}$ is the event horizon, and
if $\Lambda>0$ then the largest root is the cosmological horizon and
therefore the second largest root is the event horizon.

The thermodynamics of the black hole is determined in the usual way using path integral
methods (Hawking 1979).  In particular, the average energy $\ave{\mathfrak{E}}$ and
entropy $\mathscr{S}$ are given by
\bea
\ave{\mathfrak{E}} = -\frac{\partial}{\partial\beta}(\ln\mathcal{Z})
\quad
\mbox{and}
\quad
\mathscr{S} = \beta\ave{\mathfrak{E}} + \ln\mathcal{Z} \, ,
\label{eands}
\eea
where $\ln\mathcal{Z}$ is the (zero-loop) partition function and
$\beta$ is the inverse temperature.  The partition function is determined
via $\ln\mathcal{Z}=-\tilde{I}[g]$ by evaluating the Euclidean action
$\tilde{I}[g]$ (in the stationary phase approximation where $g$ are
solutions to the equations of motion $\delta\smallint\tilde{I}=0$), and the
inverse temperature is determined by requiring that the Euclidean manifold
does not contain any conical singularities at $r_{+}$ where the manifold
closes up.  For the black hole solution (\ref{bhsolution}) one finds that
(Cai 2002; Cho and Neupane 2002)
\bea
\ave{\mathfrak{E}} = M
\quad
\mbox{and}
\quad
\mathscr{S} = \frac{\mathscr{A}_{D-2}r_{+}^{D-2}}{4G_{D}}
    \left[1 + \left(\frac{D-2}{D-4}\right)
    \frac{2\tilde{\alpha}k}{r_{+}^{2}}\right] \; .
\label{entropy5}
\eea
Here, $\mathscr{A}_{N-1}=2\pi^{N/2}/\Gamma(N/2)$ is the surface area of a unit
$(N-1)$-sphere.  This shows that the entropy acquires a correction due to the
presence of the GB term.  For the solution (\ref{bhsolution}), the Ricci scalar
is $\mathcal{R}=(D-2)(D-3)k/r_{+}^{2}$, and (\ref{entropy4}) reduces to
(\ref{entropy5}).  Our entropy expression is therefore in agreement with the
Euclidean expression as well.  In our derivation, however, the entropy
(\ref{entropy4}) automatically satisfies the first law (\ref{firstlaw5}).

An interesting consequence of the GB term is that it is possible for black
holes to have negative entropies when $2\alpha\mathcal{R}<1$.  This was first
discovered by Cveti$\check{\mbox{c}}$ \emph{et al} (2002) and later confirmed
by Clunan \emph{et al} (2004).  For non-rotating horizons, the first law
(\ref{firstlaw5}) implies that the energy is also negative; this is not
surprising, as negative-energy solutions are possible when $\Lambda<0$
(Horowitz and Myers 1999).  We will now proceed to show that the presence
of the GB term also has consequences for the area-increase law during the
merging of two black holes.

\subsection{Decrease of black-hole entropy in EGB theory}

For any EGB black hole in $D$ dimensions, the Ricci scalar $\mathcal{R}$ integrates
to a constant over $\mathbb{S}^{D-2}$.  It is not too surprising then, that the area
always increases in any physical process involving just one black hole with an entropy
of the form (\ref{entropy4}) (Jacobson \emph{et al} 1995).  However, this will not be
the case for a system with \emph{dynamical topologies} such as black-hole mergers (Witt
2007).  This is a form of topology change, which for a space with a degenerate
metric is unavoidable even in classical general relativity (Horowitz 1991).  This fact
is relevant to the current problem because the entropy expression (\ref{entropy4})
holds for Killing horizons and WIHs, both of which are null surfaces on which the
induced metrics are degenerate.

As an example, let us consider the merging of two black holes -- one with mass $m_{1}$
and entropy $\mathcal{S}_{1}=[\mathcal{A}_{1}+2\alpha X(\mathbb{S}_{1})]/4G_{D}$, the
other with mass $m_{2}$ and entropy
$\mathcal{S}_{2}=[\mathcal{A}_{2}+2\alpha X(\mathbb{S}_{2})]/4G_{D}$.  Here, we have
defined the surface area $\mathcal{A}=\oint_{\mathbb{S}^{D-2}}\bm{\tilde{\epsilon}}$
and the correction term
$X(\mathbb{S})=\oint_{\mathbb{S}^{D-2}}\bm{\tilde{\epsilon}}\mathcal{R}$.  Before the
black holes merge, the total entropy is
\bea
\mathcal{S} &=& \mathcal{S}_{1} + \mathcal{S}_{2}\nonumber\\
            &=& \frac{1}{4G_{D}}[\mathcal{A}_{1} + \mathcal{A}_{2}
                + 2\alpha(X(\mathbb{S}_{1}) + X(\mathbb{S}_{2}))] \; .
\label{entropy6}
\eea
After the black holes merge, the total entropy of the resulting black hole is
\bea
\mathcal{S}^{\prime} = \frac{1}{4G_{D}}[\mathcal{A}^{\prime}
                       + 2\alpha X(\mathbb{S}^{\prime})] \; .
\label{entropy7}
\eea
The expressions (\ref{entropy6}) and (\ref{entropy7}) imply that
$\mathcal{S}^{\prime}>\mathcal{S}$ if and only if
\bea
\alpha < \frac{-(\mathcal{A}_{1} + \mathcal{A}_{2} - \mathcal{A}^{\prime})}
         {2[X(\mathbb{S}_{1}) + X(\mathbb{S}_{2}) - X(\mathbb{S}^{\prime})]} \; .
\label{bound1}
\eea
Without knowing the specific details of the black holes in question, nothing further
can be said about $\mathcal{S}$ and $\mathcal{S}^{\prime}$, or about the upper bound
(\ref{bound1}).  Let us therefore consider for concreteness the simplest case -- the
merging of two Schwarzschild black holes in four-dimensional flat spacetime.  This is
a particularly special case as the topologies are much more restricted than could be
hoped for.  First, the GB theorem [see e.g. (Hatfield 1992)] relates the correction
term to the Euler characteristic $\chi(\mathbb{S})$ via
\bea
X(\mathbb{S}) = \oint_{\mathbb{S}^{2}}\bm{\tilde{\epsilon}}\mathcal{R}
              = 4\pi\chi(\mathbb{S}) \; .
\eea
Then the Hawking topology theorem (Hawking 1972) restricts the horizon cross sections
to be two-spheres for which $\chi(\mathbb{S})=2$.  For a Schwarzschild black hole the
correction term is therefore $X(\mathbb{S})=8\pi$.  Furthermore, the surface area of a
Schwarzschild black hole is related to its mass via
\bea
\mathcal{A} = 16\pi m^{2} \, ,
\eea
whence the surface areas of the initial and final black-hole states are
\bea
\mathcal{A}_{1} = 16\pi m_{1}^{2} \, ,
\quad
\mathcal{A}_{2} = 16\pi m_{2}^{2} \, ,
\quad
\mbox{and}
\quad
\mathcal{A}^{\prime} = 16\pi(m_{1} + m_{2} - \gamma)^{2} \; .
\label{areas}
\eea
Here, a small mass parameter $\gamma\geq0$ for the surface area of the final
black-hole state has been included.  This parameter corresponds to any mass
that may be carried away by gravitational radiation during merging.  With
these expressions for the areas, the bound (\ref{bound1}) in terms of the
masses becomes
\bea
\alpha < 2m_{1}m_{2} - \gamma[2(m_{1} + m_{2}) - \gamma] \; .
\label{bound2}
\eea
Therefore the second law will be violated if $\alpha$ is greater than twice the
product of the masses of the black holes before merging minus a correction due to
gravitational radiation.

To summarize, the validity of the second law of black-hole mechanics was examined
for a physical process in which the topology is not constant.  As was shown, the
correction term appearing in the entropy (\ref{entropy5}) can lead to a violation
of the second law for certain values of the GB parameter during the merging of two
black holes.  The calculation was done for two Schwarzschild black holes in
four-dimensional flat spacetime.  However, a similar bound to (\ref{bound2}) may
presumably be derived for specific solutions in higher dimensions as well [although
in this case the topologies are not as severely restricted as they are in four
dimensions, even for Einstein gravity with $\Lambda=0$ (Helfgott \emph{et al} 2006;
Galloway 2006; Galloway and Schoen 2006)].  Incidently, the result obtained here
shows that the presence of the GB term in the action for gravity can have nontrivial
physical effects even in four dimensions, when the term is a topological invariant
of the manifold.  This is in sharp contrast to the commonly held belief within the
literature that the term is only significant in spacetimes with dimension $D\geq5$.

\section{Prospects}

\noindent{\emph{``Try to see through fainted views.  As reality disappears
in a haze.  A journey between eternal walls.  The senses unfold before my
eyes.''}} $\sim$ T G Fischer

\subsection{Summary}

In this thesis we presented two extensions of the IH framework, first to EM-CS theory
and then to EGB theory in $D$ dimensions.  In particular, we derived the local version
of the zeroth and first laws of black-hole mechanics for general WIHs on the phase space
of the corresponding theories.  In addition, for EM-CS theory in five dimensions and for
EM theory in four dimensions we derived the conditions that are required by supersymmetry.
We then turned to  EGB theory, for which we showed that the quasilocal entropy is in exact
agreement with the expressions that are obtained by the Euclidean and Noether charge
methods.  Finally, we showed that the GB term can have physical consequences in four
dimensions even though it is a topological invariant and does not contribute to the
equations of motion.

As was stated in $\S 1.2$, our intention was to employ the IH framework with suitable
extensions to higher dimensions in order to determine the generic properties of black
holes in string-inspired gravity models.  A summary of the main five results are given
in $\S 1.3$.

\subsection{Classical applications to EM-CS and EGB theories}

There are a number of classical applications of IHs to EM-CS and EGB black holes that
can be explored.  Here we briefly discuss four problems that are worth investigating.

Let us first consider applications to EM-CS theory, and in particular to the corresponding
supersymmetric black holes.

\begin{itemize}

\item
\emph{BPS bounds.}
The general method for deriving the BPS bound for stationary spacetimes is to construct
an expression for the energy of the spacetime using spinor identities and the Einstein
field equations.  This method was pioneered by Witten (1981) and Nester (1981) to prove
the positive energy theorem.  The method was applied in four dimensions (Gibbons and Hull
1982) and in five dimensions (Gibbons \emph{et al} 1994) to calculate the BPS bounds for
the corresponding spacetimes.  How can one derive these bounds for IHs?  The bounds are
saturated when the spinors are supercovariantly constant, which is associated with
extremality.  This suggests that the extremality condition (\ref{characterization}) can
be used for IHs.  This is straight-forward to do for undistorted horizons.  Let us
consider the four-dimensional EM theory for illustration.  Here the contraction
$T_{ab}\ell^{a}n^{b}$ is the square of the electric flux $E_{\perp}$ crossing the
surface (Ashtekar \emph{et al} 2000c).  For any IH this quantity is constant over
$S^{2}$ and can therefore be moved outside the integral.  The result can be used to
relate the charge $\mathcal{Q}$ on the horizon to its surface area $\mathcal{A}$ via
$\mathcal{Q}=E_{\perp}\mathcal{A}/(4\pi)$.  For the RN solution one finds that
$\eta=\mathcal{Q}^{2}/R^{2}-1\leq0$ with $R=\sqrt{\mathcal{A}/(4\pi)}$ the areal radius
(Booth and Fairhurst 2007b).  When the surface gravity vanishes $\eta=0$ and
$\mathcal{Q}=R=M$ with $M$ the mass.  This is the condition for supersymmetry in four
dimensions.  The situation is not as obvious for distorted horizons in five dimensions.
This is because the contraction $T_{ab}\ell^{a}n^{b}$, which for EM-CS theory is again
the square of the electric flux, may not be constant over the horizon cross sections
in general.  However, for the BMPV black hole in particular we know that the cross
sections are $S^{3}$ which have constant curvature, and therefore $E_{\perp}^{2}$ is
constant on $\Delta$.  From here, a charge-areal radius relation follows along the
same lines as the derivation that was outlined above for the RN black hole in four
dimensions.

\item
\emph{Supersymmetry and horizon geometries.}
It was shown (Lewandowski and Pawlowski 2003) that the IH constraints for extremal
IHs of four-dimensional EM theory are satisfied iff the intrinsic geometry of the
horizon coincides with that of the extremal Kerr-Newman (KN) solution.  An extension
of that analysis to IHs of five-dimensional EM-CS theory would be of interest,
particularly because it would provide a method for deriving the geometries of the
corresponding extremal IHs.  This would complement a recently developed method
(Astefanesei and Yavartanoo 2007; Kunduri \emph{et al} 2007b; Kunduri and Lucietti 2007)
for deriving the \emph{near-horizon} geometries of extremal black holes.  While
speculating on the local uniqueness theorems in five dimensions we need to keep in mind
that black holes in five dimensions are much less constrained than in four dimensions,
mainly because in five dimension there are two possible topologies ($S^{3}$ and
$S^{1}\times S^{2}$), and also because there are two independent rotation parameters.
As a consequence of this richer structure, it is possible that two distinct black
holes in five dimensions can have the same asymptotic charges (Emparan and Reall 2002).
This is a striking example of black-hole nonuniqueness in higher dimensions.
Nevertheless, uniqueness has been established for supersymmetric black holes in five
dimensions (Reall 2003).  Therefore it is expected that the five-dimensional analogues
of the local uniqueness theorem of (Lewandowski and Pawlowski 2003) do exist, but for
SIHs.  We also note that, while supersymmetry constrains the geometry (i.e. connection
one-form), the dominant energy condition and the Einstein field equations are still
required to constrain the topology.  Therefore we expect that there should be a unique
horizon geometry for a given topology.  For example, if the topology of a SIH is
$S^{1}\times S^{2}$, then the geometry should coincide uniquely with the induced
metric and vector potential of the EEMR black ring solution.  We also expect that,
if the topology is $S^{3}$, then the geometry should coincide uniquely with that of
the BMPV black hole \emph{in general}, and the extremal RN black hole as a limiting
case when the angular momentum of the Maxwell fields vanishes.  It would be of
considerable interest to try solving the IH constraints for a SIH when
$T_{ab}\ell^{a}n^{b}=0$ (at the horizon); the resulting geometry would provide the
first explicit solution of a supersymmetric black hole with toroidal topology.

\end{itemize}

Now let us consider applications to EGB theory.

\begin{itemize}

\item
\emph{Rotation.}
One of the main assumptions that we made in our calculations was that the horizons are
non-rotating.  It would be interesting to extend the IH phase space to include rotation
by relaxing the condition $\tomega=0$.

\item
\emph{Torsion.}
The formalism presented here can be further extended by including torsion.  Recall that in
$\S 5.2$ we assumed $T^{I}=0$ directly, which became crucial when we derived the pull-back
to $\Delta$ of the connection.  However, as the equation of motion (\ref{eom4}) for $A$
indicates, the torsion-free condition is not imposed in $D\geq5$ dimensions.  If the
torsion is non-zero then the pull-back to $\Delta$ of $A$ will not be given by
(\ref{pullbackofa}).  In order to derive the modified pull-back of $A$ in the presence of
torsion we would need to find $\nabla_{\! \underleftarrow{a}}e_{\phantom{a}I}^{b}$
explicitly.  In addition, the Raychaudhuri equation would be different as well, and so
the boundary conditions would require a more careful analysis.  The effects of torsion on
IHs should therefore lead to some interesting consequences.  This would be a particularly
interesting project to work out in five dimensions, for which a solution has recently been
found that describes a supersymmetric black hole (Canfora \emph{et al} 2008).  More
interestingly, there is also a solution of the equations with non-zero torsion that
describes a constant-curvature black hole with entropy that is proportional to the surface
area of the \emph{inner} horizon rather than the event horizon (Ba$\tilde{\mbox{n}}$ados
1998).  This curious interchange of thermodynamic parameters, namely the outer and inner
horizons, may be a consequence of the torsion that is present in the equations of motion.
The IH framework could be employed in order to test this hypothesis.

\end{itemize}

There are many more avenues to explore.  We hope that others will consider some of them.

\section*{Acknowledgements}

I would like to thank my Ph.D advisor Ivan Booth for collaborations,
discussions and most importantly for supporting my ideas.  I would
also like to thank Abhay Ashtekar, Dumitru Astefanesei, Kirill
Krasnov and Don Witt for correspondence related to some of the work
presented in this thesis, and especially Kirill for taking time to
answer my naive questions about quantum gravity.

This work was funded by the Natural Sciences and Engineering Research
Council of Canada (NSERC) and by funds from the School of Graduate Studies
(SGS).  Participation at conferences was supported in part by NSERC and
by travel funds received from SGS and the Department of Physics.
Participation at the Topical Workshop on Black Holes and Fundamental
Theory at Penn State University was funded by NSERC.

My deepest gratitude goes to my daughters Kristijana and Veronika for
always reminding me that there is more to life than physics, to my wife
Jana for keeping me focused on getting my work done quickly, and to my
parents Maria and $\check{\mbox{S}}$tefan for continued encouragement.

\appendix

\section{Black Hole Mechanics: An Overview}

\subsection{Thermodynamics}

The study of macroscopic properties of materials without knowing their internal
structure is the science of thermodynamics.  This is the branch of science concerned
with the dynamics of materials where thermal effects are important.  Because no reference
is made to the internal structure, the formalism involved is very general and therefore
very powerful.  Let us quickly review the four laws of thermodynamics (Reif 1965; Poisson
2000); this will make the connection between the laws of black-hole mechanics and the
four laws of thermodynamics apparent.

Let us now state the four laws of thermodynamics and discuss some of their physical
consequences.

\begin{itemize}

\item
\emph{Zeroth law.}
If two systems are each in thermal equilibrium with a third system, then they are in
thermal equilibrium with each other.  This implies that the temperature remains constant
throughout the systems that are in thermal equilibrium with each other.

\item
\emph{First law.}
The internal energy $\mathscr{U}$ of a system that interacts with its surroundings will
undergo a change given by
\bea
d\mathscr{U} = d\mathscr{Q} - d\mathscr{W},
\eea
where $d\mathscr{Q}$ is the amount of heat absorbed by the system and $d\mathscr{W}$ is
the amount of work done by the system during its change of macrostate.  This is really
just the statement of conservation of energy.  It implies that a gain of heat to a system
can do physical work on its surroundings.

\item
\emph{Second law.}
Heat flows spontaneously from higher temperatures to lower temperatures.  This means
that: (i) the spontaneous tendancy of a system to go toward thermal equilibrium cannot
be reversed without changing some organized energy (work) into some disorganized energy
(heat); (ii) it is not possible to convert heat from a hot reservoir into work in a
cyclic process without transferring some heat to a colder reservoir; (iii) the change
in entropy $d\mathscr{S}=d\mathscr{Q}/T$ (with $T$ the temperature) of a system and its
surroundings is positive and approaches zero for any process that approaches
reversibility.

\item
\emph{Third law.}
The difference in entropy $\delta\mathscr{S}$ between states connected by a reversible
process goes to zero as the temperature $T$ goes to absolute zero.  Unlike the other
laws which are based on classical considerations, the third law is a consequence of
quantum mechanics.  The third law implies that a system at absolute zero will drop to
its lowest quantum state and thus become completely ordered.

\end{itemize}

The first law expresses the change in internal energy of a system in terms of inexact
differentials.  This means that, unlike the difference $d\mathscr{Q}-d\mathscr{W}$,
seperately $d\mathscr{Q}$ and $d\mathscr{W}$ depend on the path that is taken from the
initial state to final state.  The first law can be expressed in terms of exact
differentials though.  First, we note that in a quasi-static process where the volume
of a fluid changes but the pressure remains approximately unchanged, the physical work
done by the system is given by $d\mathscr{W}=Pd\mathscr{V}$.  Then, modulo the second
law we can write the first law in the more familiar form:
\bea
d\mathscr{U} = Td\mathscr{S} - Pd\mathscr{V}.
\label{firstlaw1}
\eea
This is the form of the first law that appears in most references on gravitational
physics that discuss the laws of black-hole mechanics.

\subsection{Black-hole mechanics: global equilibrium}

Now we will review the corresponding laws of black hole mechanics.  A beautiful
introduction is given in the book by Poisson (2004).  The laws of black-hole mechanics
were first formulated for globally stationary spacetimes in four-dimensional Einstein
gravity (Bardeen \emph{et al} 1973), and later extended using covariant phase space
methods to $D$-dimensional spacetimes in arbitrary diffeomorphism-invariant theories
(Wald 1993; Iyer and Wald 1994; Jacobson \emph{et al} 1994).  This seminal work has
revealed, among other properties of black holes, that the area-entropy relation will
only be modified in cases when gravity is supplemented with nonminimally coupled
matter or higher-curvature interactions.  This is a consequence of the fact that
such terms modify the gravitational surface term in the symplectic structure.  Such
terms naturally arise in the effective actions of superstrings and supergravity.
For now we will restrict our review to $D$-dimensional black holes of EM theory.

Let us first state some facts about Killing horizons.  We shall consider the black
hole region
\bea
B = \mathcal{M}-J^{-}(\mathcal{I}^{+})
\eea
of a spacetime manifold $\mathcal{M}$; i.e. a region of spacetime that excludes all
events that belong to the causal past of future null infinity.  The event horizon is
the boundary $\partial B$ of the region $B$ of spacetime. If the black hole is stationary,
then the event horizon coincides with the Killing horizon -- a hypersurface at which the
vector
\bea
\zeta^{a} = t^{a} + \sum_{\iota}^{\lfloor (D-1)/2 \rfloor}\widetilde{\Omega}_{\iota}m_{\iota}^{a}
\label{killingfield}
\eea
is null.  Here, $t^{a}$ is a timelike Killing vector, $m_{\iota}^{a}$ are
$\lfloor (D-1)/2 \rfloor$ rotational spacelike Killing vectors and the coefficients
$\widetilde{\Omega}_{\iota}$ are the angular velocities of the black hole.  The vector
$\zeta^{a}$ is tangent to the null generators of the Killing horizon and therefore satisfies
the geodesic equation
\bea
\zeta^{a}\nabla_{a}\zeta^{b} = \kappa\zeta^{b} \; .
\eea
This defines the surface gravity $\kappa$ of the black hole.  Equivalent definitions are
given by
\bea
2\kappa\zeta_{a} = (-\zeta_{b}\zeta^{b})_{;a}
\quad
\mbox{and}
\quad
\kappa^{2} = -\frac{1}{2}\zeta_{a;b}\zeta^{a;b} \; .
\eea
For the Schwarzschild solution (\ref{schwarzschild}) it can be verified by
direct computation that $\kappa=1/(4M_{\odot})$.

We now state the four laws of black-hole mechanics:

\begin{itemize}

\item
\emph{Zeroth law.}
The surface gravity $\kappa$ is constant over the entire event horizon.

\item
\emph{First law.}
For a stationary black hole with energy $\mathfrak{E}$, surface area
$\mathscr{A}$, electric charge $\mathfrak{Q}$ and angular momenta
$\mathfrak{J}_{\iota}$, the change in mass during a quasi-static process
is given by
\bea
\delta\mathfrak{E} = \frac{\kappa}{\kappa_{D}}\delta\mathscr{A}
                     + \widetilde{\Phi}\delta\mathfrak{Q}
                     + \sum_{\iota=1}^{\lfloor (D-1)/2 \rfloor}
                       \widetilde{\Omega}_{\iota}\delta\mathfrak{J}_{\iota} \, ,
\label{firstlaw2}
\eea
where $\widetilde{\Phi}=-A_{a}\zeta^{a}|_{r=r_{+}}$ is the electric potential at the
horizon ($r=r_{+}$) and $A_{a}$ is the vector potential.

\item
\emph{Second law.}
The surface area $\mathscr{A}$ can never decrease in a physical process if the
stress-energy tensor $T_{ab}$ satisfies the dominant energy condition
$T_{ab}\zeta^{a}\zeta^{b}\geq0$.

\item
\emph{Third law.}
The surface gravity cannot be reduced to zero by any physical process in a finite
period of time.

\end{itemize}

The laws of black-hole mechanics are very similar to the laws of thermodynamics.  If one
makes the identification $\mathscr{U}=\mathfrak{E}$, then the first law of thermodynamics
(\ref{firstlaw1}) and the first law of black-hole mechanics (\ref{firstlaw2}) require that
$\mathscr{S}\propto\mathscr{A}k_{\rm B}/l_{\rm P}^{2}$ (with physical constants restored for
the moment) and $\kappa\propto T$.  However, the latter does not seem to make sense from a
classical point of view, because a black hole presumably has zero temperature which would
imply that the entropy is infinite.  Is the similarity just a mathematical coincidence, or is
nature telling us something deep about the interplay between gravitational phenomena and
quantum mechanics?  Indeed, the identification
$\mathscr{S}\propto\mathscr{A}k_{\rm B}/l_{\rm P}^{2}$ requires the presence of $c$ and
$\hbar$ in order for the entropy to be dimensionless.
%The proportionality may be written
%as an equality up to a single dimensionless parameter $z$ as
%$\mathscr{S}_{\rm BH}=z\mathscr{A}k_{\rm B}/l_{\rm P}^{2}$.
It was Bekenstein (Bekenstein 1973; Bekenstein 1974) who first recognized the physical
significance of this similarity.  Bekenstein also recognized that the irreversable process
of dropping matter into a black hole leads to the \emph{generalized} second law of thermodynamics:
\bea
\delta\mathscr{S}_{\rm Universe} + \delta\mathscr{S}_{\rm BH} \geq 0 \; .
\eea
Using a semiclassical approach, Hawking (1975) then fixed the free parameter to $1/4$ and
the temperature to $T=\kappa/2\pi$.  Thus a black hole is not eternal, but rather has
a thermodynamical temperature and radiates.  (It should be noted that the final state of
this process is unknown, because the temperature goes to infinity as the surface area
decreases.  One of the many goals of all theories of quantum gravity is to give a detailed
first-principles description of the evapouration process and to determine what the final
state of a black hole should be.)  Therefore, the laws of black-hole mechanics \emph{are}
the laws of thermodynamics, applied to an object of special character.

\section{Differential forms}

%\addcontentsline{toc}{chapter}{Appendix: Differential forms}

\subsection{Definitions}

\noindent{\bf Definition B.I:}
\emph{A differential $p$-form is a totally antisymmentric tensor of type $(0, p)$, i.e.
$\bm{\omega}_{a_{1}\cdots a_{p}}$ is a $p$-form if
\be
\bm{\omega}_{a_{1}\cdots a_{p}} = \bm{\omega}_{[a_{1}\cdots a_{p}]}.
\ee}

Denote the vector space of $p$-forms at a point $x$ by $\Lambda_{x}^{p}$ and the
collection of $p$-form fields by $\Lambda^{p}$.  Taking the outer product of a
$p$-form $\bm{\omega}_{a_{1}\cdots a_{p}}$ and a $q$-form $\bm{\mu}_{b_{1}\cdots b_{q}}$,
results in a tensor of type $(0,p+q)$, which will not be a $(p+q)$-form since this
tensor is not generally antisymmetric.

\noindent{\bf Definition B.II:}
\emph{The wedge product on an $m$-dimensional manifold $\mathcal{M}$ is a map
$\map{\wedge}{\Lambda_{x}^{p}\times\Lambda_{x}^{q}}{\Lambda_{x}^{p+q}}$
such that the tensor product
\be
(\bm{\omega} \otimes \bm{\mu})_{a_{1}\cdots a_{p}b_{1}\cdots
b_{q}}
\ee
is totally antisymmetric.}

Note that this tensor is zero if $p+q > m$.  Thus the wedge product of two one-forms
on $\mathbb{R}^{m\geq2}$ is $\bm{\omega}\wedge\bm{\mu}=-\bm{\mu} \wedge\bm{\omega}$.
Now define the vector space of all differential forms at $x$ to be the direct sum of
the $\Lambda_{x}^{p}$ such that
\be
\Lambda_{x} = \bigoplus_{p=0}^{n}\Lambda_{x}^{p}.
\ee
The map $\map{\wedge}{\Lambda_{x}^{p} \times \Lambda_{x}^{q}}{\Lambda_{x}^{p+q}}$
gives $\Lambda_{x}$ the structure of a Grassmann algebra over the vector
space of one-forms.

\noindent{\bf Definition B.III:}
\emph{The exterior derivative on an $m$-dimensional manifold $\mathcal{M}$ is a map
from the space of $p$-forms to the space of $(p+1)$-forms:
\be
\map{d}{\Lambda^{p}}{\Lambda^{p+1}},
\ee
together with the following properties:
\begin{enumerate}
\item
$d(\bm{\omega}+\bm{\mu})=d\bm{\omega}+\bm{d\mu}$;
\item
$d(c\bm{\omega})=cd\bm{\omega}$;
\item
$d(\bm{\omega}\wedge\bm{\lambda})=d\bm{\omega}\wedge\bm{\lambda}
+(-1)^{p}\bm{\omega}\wedge d\bm{\lambda}$;
\item
$d(d\bm{\omega})=0$;
\end{enumerate}
$\forall$ $\bm{\omega},\bm{\mu}\in\Lambda^{p}(\mathcal{M})$,
$\bm{\lambda}\in\Lambda^{q}(\mathcal{M})$, and $c\in\mathbb{R}$.}

The last property can be easily shown as follows.  Consider the $p$-form
$\bm{\omega}$ such that
\be
\bm{\omega} = \frac{1}{p!}\omega_{a_{1}\cdots a_{p}}
                      dx^{a_{1}}\wedge\cdots\wedge dx^{a_{p}}.
\ee
The exterior derivative acting on $\bm{\omega}$ is given by
\be
d\bm{\omega} = \frac{1}{p!}\partial_{\mu}\omega_{a_{1}\cdots a_{p}}
                       dx^{\mu}\wedge dx^{a_{1}}\wedge\cdots\wedge dx^{a_{p}},
\ee
and the second exterior derivative is
\be
dd\bm{\omega} = \frac{1}{p!}\partial_{\nu}\partial_{\mu}\omega_{a_{1}\cdots a_{p}}
                        dx^{\nu}\wedge dx^{\mu}\wedge dx^{a_{1}}\wedge\cdots\wedge dx^{a_{p}}.
\ee
Since the functions $\omega_{a_{1}\cdots a_{p}}$ are by definition smooth, the partial
derivatives acting on them commute and the operator $\partial_{\mu}\partial_{\nu}$ is
symmetric.  The two-form $dx^{\mu}\wedge dx^{\nu}$, however, is antisymmetric.  Thus we
have $dd\bm{\omega}=0$.  This important property is often written simply $d^{2}=0$.

In general, if $\bm{\alpha}\in\Lambda^{p+1}(\mathcal{M})$ and
$\bm{\beta}\in\Lambda^{p}(\mathcal{M})$ then: $\bm{\alpha}$ is called an exact form
iff $d\bm{\alpha}=0$; $\bm{\alpha}$ is called a closed form iff
$\bm{\alpha}=d\bm{\beta}$.  By the last property $d^{2}=0$ of the exterior derivative,
a closed form is automatically an exact form as well.  However, the converse is not
true, and the study of this is called DeRham cohomology.

\noindent{\bf Definition B.IV:}
\emph{The interior product of a $p$-form $\bm{\omega}$ with vector field $X$
on an $m$-dimensional manifold $\mathcal{M}$ is a map
$\map{i_{X}}{\Lambda^{p}(\mathcal{M})}{\Lambda^{p-1}(\mathcal{M})}$ such that
\be
i_{X}\bm{\omega}(X_{1},\ldots,X_{p-1}) = \bm{\omega}(X,X_{1},\ldots,X_{p-1})
\ee
together with the ``anti-derivation'' property with respect to the wedge product
\be
i_{X}(\bm{\omega}\wedge\bm{\eta}) = (i_{X}\bm{\omega})\wedge\bm{\eta}
+ (-1)^{p}\bm{\omega}\wedge(i_{X}\bm{\eta})
\ee
where $\bm{\omega}\in\Lambda^{p}(\mathcal{M})$ and
$\bm{\eta}\in\Lambda^{q}(\mathcal{M})$.  If $\bm{\omega}\in\Lambda^{0}(\mathcal{M})$
then $i_{X}\bm{\omega}=0$.}

The interior product is also called the contraction, and is also denoted
$X \lrcorner \bm{\omega}$.  As an example, consider
$\bm{\omega}\in\Lambda^{2}(\mathcal{M})$ and
$Z=X^{\alpha}(\partial/\partial x^{\alpha})+Y^{\alpha}(\partial/\partial y^{\alpha})$
a vector field on $\mathcal{M}$.  The interior product of $\bm{\omega}$ and $Z$ is
given by
\be
Z \lrcorner \bm{\omega} = -Y_{\alpha}dx^{\alpha} + X_{\alpha}dy^{\alpha}.
\ee

\noindent{\bf Definition B.V:}
\emph{The Lie derivative of a $p$-form $\bm{\omega}$ with respect to a vector field $X$ is given by}
\be
\pounds_{X}\bm{\omega} = X \lrcorner d\bm{\omega} + d(X \lrcorner \bm{\omega}) \; .
\ee

\noindent{\bf Definition B.VI:}
\emph{The Hodge star operator on an $m$-dimensional manifold $\mathcal{M}$ is a
linear map $\map{*}{\Lambda^{p}(\mathcal{M})}{\Lambda^{m-p}(\mathcal{M})}$ given
by
\be
*\left(\bm{e}_{a_{1}}\wedge\cdots\wedge\bm{e}_{a_{p}}\right)
= \frac{1}{(m-p)!}\epsilon_{a_{1}\cdots a_{p}}^{\phantom{a_{1}\cdots a_{p}}a_{p+1}\cdots a_{m}}
  \bm{e}_{a_{p+1}}\wedge\cdots\wedge\bm{e}_{a_{m}},
\ee
where $\lbrace\bm{e_{a}}\rbrace_{a=1}^{m}$ is a positively-oriented set of one-forms
on some chart of $\mathcal{M}$.}

As a simple example, consider a two-form on $\mathbb{R}^{3}$.  Choosing a
basis $\lbrace\bm{e}_{1}\wedge\bm{e}_{2},\bm{e}_{1}\wedge
\bm{e}_{3},\bm{e}_{2}\wedge\bm{e}_{3}\rbrace$, the definition gives
$*\left(\bm{e}_{i}\wedge\bm{e}_{j}\right)=\epsilon_{ij}^{\phantom{aa}k}\bm{e}_{k}$,
or $*\left(\bm{e}_{1}\wedge\bm{e}_{2}\right)=\bm{e}_{3}$,
$*\left(\bm{e}_{1}\wedge\bm{e}_{3}\right)=-\bm{e}_{2}$ and
$*\left(\bm{e}_{2}\wedge\bm{e}_{3}\right)=\bm{e}_{1}$.

\end{document}